\documentclass[longauth]{aa}
\usepackage{graphicx}
\usepackage{newtxtext, newtxmath}
\usepackage{hyperref}
\usepackage{overpic} 
\usepackage{natbib}
\usepackage{booktabs}
\usepackage{float}
\usepackage{multirow}
\usepackage{amstext}

\newcommand{\dd}{\mathrm{d}}
\newcommand{\degree}{$^{\circ}$}
\newcommand{\simm}{\sim\hspace{-3pt}}
\newcommand{\eg}{\textit{e.g.}}
\newcommand{\ie}{\textit{i.e.}}

\defcitealias{perotto_calibration_2020}{P20}
\newcommand{\calib}{\citetalias{perotto_calibration_2020}}
\defcitealias{arnaud_universal_2010}{A10}
\newcommand{\upp}{\citetalias{arnaud_universal_2010}}
\defcitealias{hasselfield_atacama_2013}{H13}
\newcommand{\act}{\citetalias{hasselfield_atacama_2013}}

\begin{document} 

\title{Exploiting NIKA2/XMM-\textit{Newton} imaging synergy for intermediate-mass high-$z$ galaxy clusters within the NIKA2 SZ Large Program}
\subtitle{Observations of ACT-CL~J0215.4+0030 at $z \sim 0.9$}

\titlerunning{NIKA2 observations of an intermediate-mass cluster}

\author{  F.~K\'eruzor\'e        \inst{\ref{LPSC}}
     \and F.~Mayet               \inst{\ref{LPSC}}
     \and G.W.~Pratt             \inst{\ref{CEA}}
     \and R.~Adam                \inst{\ref{LLR}}
     \and P.~Ade                 \inst{\ref{Cardiff}}
     \and P.~Andr\'e             \inst{\ref{CEA}}
     \and A.~Andrianasolo        \inst{\ref{IPAG}}
     \and M.~Arnaud              \inst{\ref{CEA}}
     \and H.~Aussel              \inst{\ref{CEA}}
     \and I.~Bartalucci          \inst{\ref{CEA}}
     \and A.~Beelen              \inst{\ref{IAS}}
     \and A.~Beno\^it            \inst{\ref{Neel}}
     \and S.~Berta               \inst{\ref{IRAMF}}
     \and O.~Bourrion            \inst{\ref{LPSC}}
     \and M.~Calvo               \inst{\ref{Neel}}
     \and A.~Catalano            \inst{\ref{LPSC}}
     \and M.~De~Petris           \inst{\ref{Roma}}
     \and F.-X.~D\'esert         \inst{\ref{IPAG}}
     \and S.~Doyle               \inst{\ref{Cardiff}}
     \and E.~F.~C.~Driessen      \inst{\ref{IRAMF}}
     \and A.~Gomez               \inst{\ref{CAB}} 
     \and J.~Goupy               \inst{\ref{Neel}}
     \and C.~Kramer              \inst{\ref{IRAMF},\ref{IRAME}}
     \and B.~Ladjelate           \inst{\ref{IRAME}} 
     \and G.~Lagache             \inst{\ref{LAM}}
     \and S.~Leclercq            \inst{\ref{IRAMF}}
     \and J.-F.~Lestrade         \inst{\ref{LERMA}}
     \and J.~F.~Mac\'ias-P\'erez \inst{\ref{LPSC}}
     \and P.~Mauskopf            \inst{\ref{Cardiff},\ref{Arizona}}
     \and A.~Monfardini          \inst{\ref{Neel}}
     \and L.~Perotto             \inst{\ref{LPSC}}
     \and G.~Pisano              \inst{\ref{Cardiff}}
     \and E.~Pointecouteau       \inst{\ref{IRAP}}
     \and N.~Ponthieu            \inst{\ref{IPAG}}
     \and V.~Rev\'eret           \inst{\ref{CEA}}
     \and A.~Ritacco             \inst{\ref{IAS},\ref{ENS}}
     \and C.~Romero              \inst{\ref{Pennsylvanie}}
     \and H.~Roussel             \inst{\ref{IAP}}
     \and F.~Ruppin              \inst{\ref{MIT}}
     \and K.~Schuster            \inst{\ref{IRAMF}}
     \and S.~Shu                 \inst{\ref{IRAMF}} 
     \and A.~Sievers             \inst{\ref{IRAME}}
     \and C.~Tucker              \inst{\ref{Cardiff}}
     }
\institute{
     Univ. Grenoble Alpes, CNRS, Grenoble INP, LPSC-IN2P3, 53, avenue des Martyrs, 38000 Grenoble, France
     \label{LPSC}
     \and
     AIM, CEA, CNRS, Universit\'e Paris-Saclay, Universit\'e Paris Diderot, Sorbonne Paris Cit\'e, 91191 Gif-sur-Yvette, France
     \label{CEA}
     \and
     LLR (Laboratoire Leprince-Ringuet), CNRS, École Polytechnique, Institut Polytechnique de Paris, Palaiseau, France
     \label{LLR}
     \and
     Astronomy Instrumentation Group, University of Cardiff, UK
     \label{Cardiff}
     \and
     Univ. Grenoble Alpes, CNRS, IPAG, 38000 Grenoble, France 
     \label{IPAG}
     \and
     Institut d'Astrophysique Spatiale (IAS), CNRS and Universit\'e Paris Sud, Orsay, France
     \label{IAS}
     \and
     Institut N\'eel, CNRS and Universit\'e Grenoble Alpes, France
     \label{Neel}
     \and
     Institut de RadioAstronomie Millim\'etrique (IRAM), Grenoble, France
     \label{IRAMF}
     \and 
     Dipartimento di Fisica, Sapienza Universit\`a di Roma, Piazzale Aldo Moro 5, I-00185 Roma, Italy
     \label{Roma}
     \and
     Centro de Astrobiolog\'ia (CSIC-INTA), Torrej\'on de Ardoz, 28850 Madrid, Spain
     \label{CAB}
     \and  
     Instituto de Radioastronom\'ia Milim\'etrica (IRAM), Granada, Spain
     \label{IRAME}
     \and
     Aix Marseille Univ, CNRS, CNES, LAM (Laboratoire d'Astrophysique de Marseille), Marseille, France
     \label{LAM}
     \and 
     Observatoire de Paris, PSL university, Sorbonne Université, CNRS, LERMA, F-75014, Paris, France  
     \label{LERMA}
     \and
     School of Earth and Space Exploration and Department of Physics, Arizona State University, Tempe, AZ 85287, USA
     \label{Arizona}
     \and
     IRAP, Université de Toulouse, CNRS, CNES, UPS, Toulouse, France
     \label{IRAP}
     \and
     Département de Physique, Ecole Normale Supérieure, 24, rue Lhomond 75005 Paris, France
     \label{ENS}
     \and
     Department of Physics and Astronomy, University of Pennsylvania, 209 South 33rd Street, Philadelphia, PA, 19104, USA
     \label{Pennsylvanie}
     \and 
     Institut d'Astrophysique de Paris, CNRS (UMR7095), 98 bis boulevard Arago, 75014 Paris, France
     \label{IAP}
     \and
     Kavli Institute for Astrophysics and Space Research, Massachusetts Institute of Technology, Cambridge, MA 02139, USA
     \label{MIT}
   }


\abstract{%
High-resolution mapping of the intracluster medium (ICM) up to high redshift and down to low masses is crucial to derive accurate mass estimates of the galaxy cluster and to understand the systematic effects affecting cosmological studies based on galaxy clusters.
We present a spatially resolved Sunyaev-Zel'dovich (SZ) X-ray analysis of ACT-CL~J0215.4+0030, 
a high-redshift ($z=0.865$) galaxy cluster of intermediate mass ($M_{500} \simeq 3.5 \times 10^{14} \;\mathrm{M_\odot}$)
observed as part of the ongoing NIKA2 SZ Large Program, which is a follow-up of a representative sample of objects at $0.5 \leqslant z \leqslant 0.9$. 
In addition to the faintness and small angular size induced by its mass and redshift, the cluster is contaminated by point sources that significantly affect the SZ signal.
This is therefore an interesting case study for the most challenging sources of the NIKA2 cluster sample. 
We present the NIKA2 observations of this cluster and the resulting data.
We identified the point sources that affect the NIKA2 maps of the cluster as submillimeter galaxies (SMG) with counterparts in catalogs of sources constructed by the SPIRE instrument on board the \textit{Herschel} observatory.
We reconstructed the ICM pressure profile by performing a joint analysis of the SZ signal and of the point-source component in the NIKA2 150~GHz map. 
This cluster is a very weak source that lies below the selection limit of the \textit{Planck} catalog.
Nonetheless, we obtained high-quality estimates of the ICM thermodynamical properties with NIKA2.
We compared the pressure profile extracted from the NIKA2 map to the pressure profile obtained from X-ray data alone by deprojecting the public XMM-\textit{Newton} observations of the cluster.
We combined the NIKA2 pressure profile with the X-ray deprojected density to extract detailed information on the ICM.
The radial distribution of its thermodynamic properties (the pressure, temperature and entropy) indicate that the cluster has a highly disturbed core.
We also computed the hydrostatic mass of the cluster, which is compatible with estimations from SZ and X-ray scaling relations.
We conclude that the NIKA2 SZ Large Program can deliver quality information on the thermodynamics of the ICM even for one of its faintest clusters after a careful treatment of the contamination by point sources.}
      
\maketitle

\section{Introduction}

As the last stage of hierarchical structure formation, galaxy clusters constitute the most massive gravitationally bound structures in the Universe.
They form through gravitational collapse of matter and accretion of local material in the density peaks at the intersection of cosmic web filaments.
As such, they are a direct tracer of the matter distribution in the Universe, and their abundance in mass and redshift constitutes an excellent probe of large-scale structure formation physics.
The latter being dominated by gravitational processes, these observables are very sensitive to the underlying cosmology, such as the initial conditions in the primordial Universe and its contents and evolution in time \citep{huterer_growth_2015}. 
Galaxy clusters can therefore be used as cosmological probes \citep[for reviews, see, \eg,][]{carlstrom_cosmology_2002,allen_cosmological_2011}.

In the past few years, large catalogs of galaxy clusters \citep[\eg,][]{planck_collaboration_planck_2016-1, bleem_sptpol_2020, adami_xxl_2018, rykoff_redmapper_2016} have been assembled and used for cosmological purposes \citep[\eg,][]{planck_collaboration_planck_2016-2,bocquet_cluster_2019,pacaud_xxl_2018,costanzi_methods_2019}.
When combined with cosmological constraints from other probes, the results of these analyses seem to exhibit a slight discrepancy with cosmological parameters inferred from the analysis of the Cosmic microwave background (CMB) power spectrum \citep[see, \eg,][]{planck_collaboration_planck_2018,salvati_constraints_2018}.
This difference might either arise from new physics or simply reflect an incomplete knowledge of the relations and tools used for the cosmological exploitation of cluster surveys \citep{ruppin_impact_2019,salvati_mass_2019,salvati_impact_2020}.
A precise knowledge of the systematic effects on the estimation of cosmological parameters from galaxy clusters surveys is therefore crucial.

One of the main limiting factors for cluster-based cosmological analyses currently comes from the relations used to infer fundamental properties of galaxy clusters from large surveys.
In particular, every cluster-based cosmological analysis strongly relies on the knowledge of the masses of clusters \citep[see, \eg,][for a review]{pratt_galaxy_2019}.
Because most of the mass budget of a cluster comes from dark matter, the total mass of a cluster is not directly observable and must be inferred either from its gravitational effect or from empirical scaling relations that link cluster masses to observables.
These observables include probes based on the thermodynamical properties of the gaseous intra-cluster medium (ICM), such as the Sunyaev-Zel'dovich (SZ) effect in millimeter wavelengths \citep[\eg,][]{planck_collaboration_planck_2011} or X-ray luminosity \citep[\eg,][]{pratt_galaxy_2009}.

The SZ effect \citep{sunyaev_observations_1972} has been proven an excellent way to detect a large number of galaxy clusters up to high redshift because it is independent of redshift.
Moreover, the integrated SZ surface brightness is a low-scatter mass proxy \citep[see, \eg,][]{pratt_galaxy_2019,planck_collaboration_planck_2011,planck_collaboration_planck_2011-1,planck_collaboration_planck_2011-2}, which means that large catalogs of SZ-detected clusters are an efficient way to probe cosmology.
The largest cluster catalog to date was obtained from \textit{Planck} observations in millimeter wavelengths \citep{planck_collaboration_planck_2016-1} and includes $\simm 1200$ clusters up to $z\simeq 0.9$, while ground-based CMB experiments such as the Atacama Cosmology Telescope \citep[ACT,][]{hilton_atacama_2018} and the South Pole Telescope \citep[SPT,][]{bleem_sptpol_2020} reach deeper in redshift and detect clusters down to lower masses.
Most scaling relations used to estimate the mass of SZ-detected clusters were calibrated on small samples of low-redshift clusters with masses inferred from X-ray observations \cite[see][]{planck_collaboration_planck_2011,arnaud_universal_2010}.
Because these clusters are located at the low end of the redshift distribution of SZ catalogs, an evolution of the scaling relations with redshift might affect the cosmological parameters estimated from these samples significantly.
In addition to the scaling relation, SZ-based cosmological analyses also rely on the knowledge of the pressure profile of galaxy clusters.
In the self-similar scenario, galaxy clusters are expected to be scaled replicas of one another.
Their pressure distribution is therefore expected to be well described by a universal pressure profile \citep{press_formation_1974,bohringer_modelling_2012}.
Similarly to the scaling relation, some of the most frequently used measurements of universal pressure profiles have been made in low-redshift clusters \citep[\eg,][hereafter \upp]{planck_collaboration_planck_2013,arnaud_universal_2010}, and a variation of this profile with redshift might also have great implications on SZ cosmology \citep{ruppin_impact_2019}.

So far, X-ray follow-ups that are deep enough to measure annular temperature profiles of $z>0.3$ SZ-discovered clusters have concentrated on representative samples of the highest-mass systems. 
They are the easiest systems to observe because their high X-ray brightness leads to good precision on the radial temperature distribution, see, for instance, \ \citet{bartalucci_resolving_2017} for a discussion of the entropy and pressure profiles of five massive ($M > 5 \times 10^{14} \,\mathrm{M}_\odot$) clusters at $z \sim 1$. 
For these objects, the ICM properties are expected to be dominated by simple shock heating and compression due to gravitational collapse.
Nonetheless, low-mass galaxy clusters are key to understanding nongravitational processes because these effects are more apparent in their shallower potentials. 
Different processes (e.g.,\ supernovae winds, active galactic nucleus (AGN) feedback, radiative cooling) affect the ICM entropy in different ways (both in terms of the level and the characteristic timescale), so that the gas history is expected to depend sensitively on their relative contribution.
This can significantly change the profile shapes \citep{pratt_gas_2010}, affect the selection function at low mass, and introduce additional scatter into the observable-mass relations. 
The evolution and scatter of thermodynamical profiles at low mass is thus a key information needed to distinguish and understand the respective role of each process.
This makes the study of distant low-mass objects essential to our understanding of galaxy clusters as a whole.

The reconstruction of the ICM properties through SZ observations faces several challenges.
One challenge is the large diversity in the SZ fluxes of clusters.
The link between a cluster mass and its integrated SZ signal implies that low-mass clusters are fainter than massive clusters.
For follow-ups of individual clusters, this means that the time needed to reach a given significance level of detection increases when the mass of clusters decreases.
Another challenge is the contamination by point sources.
Dusty galaxies and radio sources can emit in millimeter wavelengths.
Depending on their angular position with respect to the center of the cluster, they may have a strong effect on the observed 2D shape of the ICM.
The spectral signature of the SZ effect at frequencies lower that 217 GHz is a surface brightness decrement, therefore the contamination by these point sources can compensate for the SZ flux.
To remedy this, the spectra of millimetric sources can be estimated (provided data are available at other frequencies) and extrapolated to the wavelengths of interest, giving an estimate of the amplitude of the contamination \citep[see, \eg,][]{sayers_contribution_2013, adam_high_2016, ruppin_first_2018}.
Distant clusters can also be difficult to observe because their apparent angular size diminishes with redshift.
These surveys were therefore not sensitive to potential substructures in the ICM.
High angular resolution observations of high-redshift clusters may reveal significant deviations from our knowledge of nearby clusters.
The analysis presented in this paper faces these three challenges and uses NIKA2 SZ observations to overcome them.
For a review of high angular resolution SZ observations and of the associated challenges, see \citet{mroczkowski_astrophysics_2019}.

We present the study of the second observed cluster of the NIKA2 SZ Large Program, a high-resolution follow-up of 50 SZ-detected clusters at $0.5 \leqslant z \leqslant 0.9$ 
from the \textit{Planck} and ACT galaxy cluster catalogs \citep{planck_collaboration_planck_2016-1, hasselfield_atacama_2013}.
ACT-CL~J0215.4+0030 has 
one of the lowest masses and highest redshifts in the Large Program sample, with $M_{500} \simeq 3.5 \times 10^{14} \;\mathrm{M_\odot}$ and $z = 0.865$. 
It was discovered by the Atacama Cosmology Telescope with a significance level of $5.5\sigmaup$ \citep[][for the most recent data release]{menanteau_atacama_2013, hilton_atacama_2018}, and is below the detection threshold in the \textit{Planck} catalog \citep{planck_collaboration_planck_2016-1}.
The NIKA2 maps of the cluster also exhibit a strong contamination by point sources that have a strong effect on the observed SZ signal \citep{keruzore_low-mass_2020}.
This target therefore combines the challenges associated with low mass, high redshift, and point-source contamination.
The first analysis of NIKA2 SZ observations was presented in \citet{ruppin_first_2018} as a science-verification study,
targeting one of the most massive and closest clusters of the NIKA2 SZ Large Program with a long exposure time and therefore reaching a high peak signal-to-noise ratio (S/N) of $\sim 14 \sigmaup$.
By contrast, we met the challenges associated with the characterization of a
cluster observed with standard conditions for the NIKA2 SZ Large Program in the low-mass, high-redshift part of the sample.
The study of NIKA2 observations of a comparably distant and low-mass cluster has recently been presented by \citet{ricci_xxl_2020} outside the framework of the NIKA2 SZ Large Program.
We solved similar problems, and in addition, our data were strongly contaminated by a point source.
This analysis is therefore designed as a worst-case scenario for the NIKA2 SZ Large Program.

This paper is organized as follows.
In Sect.~\ref{sec:lpsz} we describe the key elements of the NIKA2 SZ Large Program.
In Sect.~\ref{sec:data} we describe the NIKA2 and XMM-\textit{Newton} data we used for this analysis, the observations and the data reduction led to the SZ and X-ray maps of the ICM.
In Sect.~\ref{sec:ps} we  identify point sources that limit the estimation of the SZ signal of the cluster.
In Sect.~\ref{sec:panco} we present the deprojection of the pressure profile of the cluster by performing a joint fit that models both the ICM seen through the SZ effect and the point sources with known positions.
In Sect.~\ref{sec:results} we present the results of this analysis through the recovered pressure profile and total integrated SZ signal, and combine this pressure profile with X-ray data to completely characterize the thermodynamics of the ICM.
We conclude and describe the implications for the NIKA2 SZ Large Program in Sect.~\ref{sec:conclu}.

Throughout this paper, we assume a flat $\Lambda$CDM cosmology with $H_0 = 70\;\mathrm{km/s/Mpc}$, $\Omega_{m,0}=0.3$, and $\Omega_{\Lambda,0}=1-\Omega_{m,0}=0.7$.
With this assumption, an angular size of 1 arcmin is subtended by a distance of 462 kpc at the redshift of ACT-CL~J0215.4+0030, that is,\ $z=0.865$.

\section{NIKA2 SZ Large Program}\label{sec:lpsz}

We present the first standard analysis of an individual cluster within the NIKA2 SZ Large Program (hereafter LPSZ).
In this section, we briefly describe this program; more detailed information can be found in \citet{mayet_cluster_2020}.

The NIKA2 camera \citep{adam_nika2_2018,bourrion_nikel_amc_2016,calvo_nika2_2016} is a dual-band millimeter camera operated at the IRAM 30-meter telescope located at Pico Veleta, Spain.
It uses kinetic inductance detectors \citep[KIDs,][]{day_broadband_2003,shu_optical_2018} to simultaneously map the sky at 150 and 260 GHz with a large field of view ($6.5'$), and a high angular resolution ($17.6"$ and $11.1"$ at 150 and 260 GHz, respectively) and sensitivity (9 and 30 $\mathrm{mJy} \cdot \mathrm{s}^{1/2}$ at 150 and 260 GHz, respectively).
The performance and detailed specifications of NIKA2 are presented in \citet[][hereafter \calib]{perotto_calibration_2020}.
NIKA2 and its pathfinder, NIKA \citep{catalano_performance_2014}, have proven to be especially well designed for SZ analyses of intermediate- to high-redshift clusters, taking full advantage of the instrument performance to extract information about the thermodynamics of the ICM \citep[see, \eg,][]{adam_first_2014,adam_pressure_2015,adam_high_2016,adam_mapping_2017-1,adam_mapping_2017,adam_substructure_2018,ricci_xxl_2020,romero_multi-instrument_2018,ruppin_non-parametric_2017,ruppin_first_2018,ruppin_unveiling_2020}.

As part of the NIKA2 guaranteed time, the NIKA2 SZ Large Program seeks to obtain a high-resolution follow-up of $\simm 50$ SZ-detected galaxy clusters from the \textit{Planck} and ACT catalogs \citep{planck_collaboration_planck_2016-1, hasselfield_atacama_2013}.
This sample covers a wide range of masses ($3 \leqslant M_{500} / 10^{14}\;\mathrm{M}_\odot \leqslant 10$) at intermediate to high redshifts ($0.5 \leqslant z \leqslant 0.9$).
The high-resolution observations of the ICM with NIKA2, combined with prior knowledge of the cluster integrated SZ fluxes from the \textit{Planck} and ACT catalogs, will allow us to achieve a precise knowledge of the dynamical state of the ICM for high-redshift clusters, with the ability to detect substructures in clusters (\eg,\ overpressure regions or merger events).
This will enable an evaluation of the mean pressure profile of galaxy clusters at higher redshift than current measurements and with a high precision.
In addition, the combination of NIKA2 SZ observations with X-ray follow-ups of clusters in the LPSZ sample will enable the computation of hydrostatic masses of the clusters, which will be used to calibrate the SZ--mass scaling relation at high redshift.
The high-redshift, high-precision measurement of these two ingredients will be used to reanalyze the large SZ cluster surveys with an evaluation of the possible redshift evolution of the relations, and with an improved control of the systematics coming from the knowledge of the dynamical state of the ICM \citep{ruppin_impact_2019,ruppin_impact_2019-1}.
The NIKA2 observations of the LPSZ sample are ongoing.

\section{Observations of the intracluster medium in ACT-CL J0215.4+0030}
\label{sec:data}

The analysis presented in this paper is based on observations of ACT-CL~J0215.4+0030 with NIKA2 and XMM-{\it Newton}.
In this section, we present the observations, the data reduction procedure and the resulting maps obtained with both instruments.

\subsection{NIKA2 data}\label{subsec:nk2data}

\subsubsection{The thermal Sunyaev-Zel'dovich effect}\label{subsec:SZ}
Millimeter observations of galaxy clusters enable the mapping of their ICM through the thermal Sunyaev-Zel'dovich effect.
It consists of the inverse Compton scattering of CMB photons on the hot electrons of the ICM, which causes a spectral distortion of the CMB.
The relative variation in CMB intensity caused by the tSZ effect is given by
\begin{equation}
    \left[\frac{\Delta I}{I_0}\right]_\mathrm{SZ} = y\, f(\nu, T_e),
    \label{eq:szi}
\end{equation}
where $I_0 = 2 (k_{\rm B} T_{\rm CMB})^3 \times (hc)^{-2}$ is the CMB specific intensity,
$T_e$ is the electron temperature of the ICM, $f(\nu, T_e)$ is the spectral dependence of the SZ effect, and $y$ is the amplitude of the distortion, referred to as the Compton parameter, which is proportional to the line-of-sight integrated pressure of the ICM,
\begin{equation}
    y = \frac{\sigma_\textsc{t}}{m_e c^2}\int P_e \, \dd l.
    \label{eq:szy}
\end{equation}
Observations of galaxy clusters through the SZ effect can therefore be used to reconstruct the pressure distribution in the ICM. 
We note that Eqs. \ref{eq:szi} and \ref{eq:szy} do not include any redshift dependence, meaning that the SZ effect is not cosmologically dimmed when high-redshift clusters are observed.
It is therefore an ideal probe to detect and study galaxy clusters on wide ranges of redshifts, as long as the angular resolution of the observations is a high enough 
\citep[for a review of SZ effect observations and of the physics involved, see, \eg, ][]{mroczkowski_astrophysics_2019}.

\subsubsection{Observations and data reduction}\label{subsubsec:obs_data_reduc}

\begin{figure*}[!thp]
    \centering
    \includegraphics[width=\linewidth]{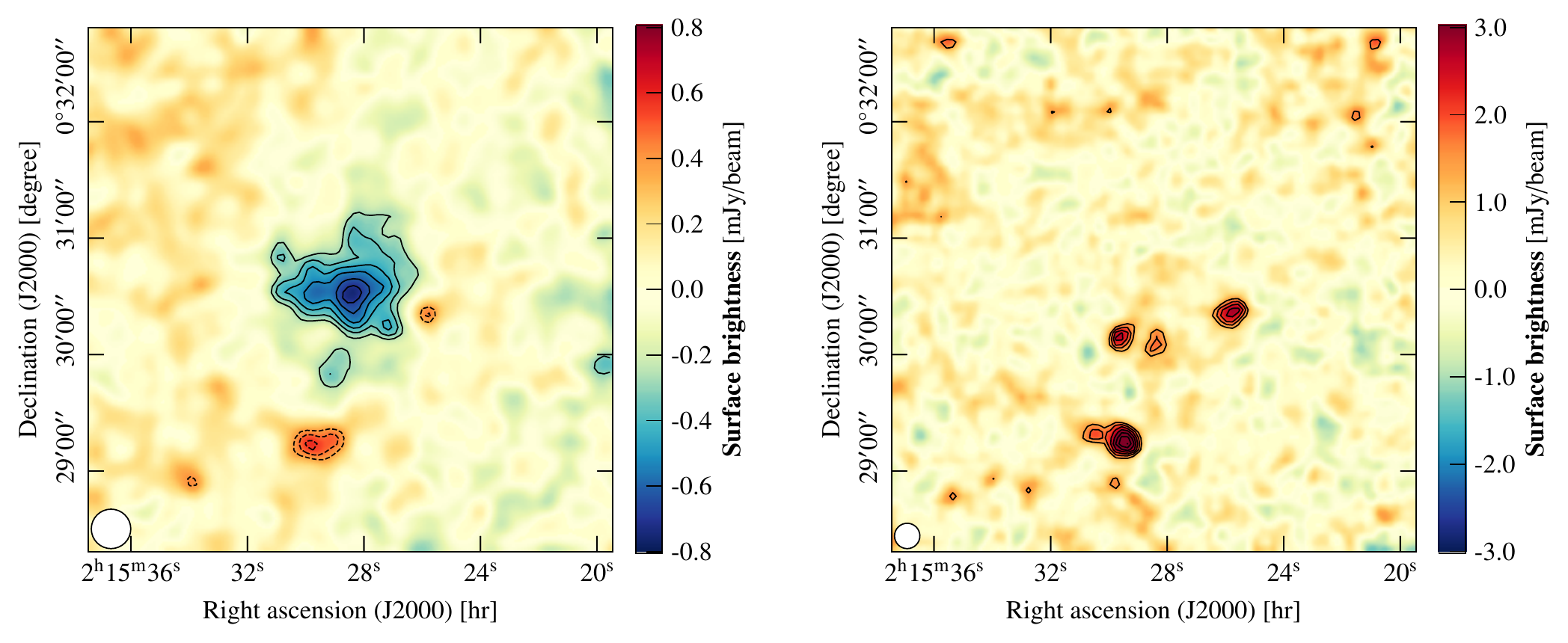}
    \caption[NIKA2 maps]{%
        NIKA2 maps of the cluster in the 150~GHz (\textit{left}) and 260~GHz (\textit{right}) bands. %
        Both maps are shown in a 4.5'$\times$4.5' square centered on the observed coordinates. %
        In both maps, contours show S/N levels starting from $\pm 3\sigmaup$ with a $1\sigmaup$ spacing. %
        The left (right) map is smoothed with a 10'' (6'') Gaussian kernel for display purposes.
        The effective FWHM for each map is represented as a white disk in the bottom left corner. %
        }
    \label{fig:nk2maps}
\end{figure*}

The NIKA2 observations used in this paper were conducted during the fourteenth run of NIKA2 (N2R14).
This run was one of those used for the performance assessment of NIKA2 and is therefore thoroughly described in \calib; we give here the key elements relevant to our observations.
The observations of ACT-CL~J0215.4+0030 were performed in five sessions from 17 to 22 January 2018.
All sessions occurred between 16:00 and 22:15 UT.
The focus of the telescope was checked at the beginning of each session.
The pointing coordinates and beam stability were checked on average every hour and were deemed stable enough that no focus corrections were needed during the observing sessions.
The cluster was observed for a total of 8.6 hours in stable conditions and with an average opacity of $\tau_{225} = 0.175 \pm 0.05$.
Because standard LPSZ conditions required nine hours of observations with an average zenith opacity at 225 GHz $\tau_{225}=0.1$, we expect the S/N in our recovered profiles to fall slightly short of our goal.

The observations were performed as a series of four raster scans of $8\times4$ arcminutes with 10 arcsecond steps between subscans, with angles of 0, 45, 90, and 135 degrees with respect to the right-ascension axis.
The center of these scans was chosen at the coordinates reported for this cluster by the first ACT SZ catalog, that is,\ (RA,~Dec)$_\text{J2000}$ = (02h15m28.8s +00\degree 30'33.0'') \citep{hasselfield_atacama_2013}.

The NIKA2 data were calibrated following the baseline procedure described in \calib.
They were then reduced using the \textit{\textup{most correlated pixels}} method from \calib : common modes are adjusted on the time-ordered information (TOIs) by groups of most-correlated detectors outside of a 2 arcminute diameter disk, and subtracted from the data as a correlated noise estimate.
The filtering of large angular scales induced by this data processing is evaluated by computing a transfer function on a simulation including synthetic signal and correlated noise.
This allows us to identify that the signal is well preserved (with a transfer function greater than 0.8) down to wavenumbers of $0.5\;{\rm arcmin}^{-1}$.
The power spectrum of the residual correlated noise is computed on a noise map obtained from the half-differences of the individual scan maps.
More details on the noise power spectrum and the transfer function are given in Appendix~\ref{ap:noisetf}.
More details on data reduction and calibration can be found in \calib, as well as in previous NIKA2 (and NIKA) SZ papers, for example, \citet{ruppin_first_2018}, \citet{adam_pressure_2015} and \citet{adam_high_2016}.

\subsubsection{NIKA2 maps of the cluster}\label{subsubsec:nk2_maps}
The resulting NIKA2 maps are presented in Fig.~\ref{fig:nk2maps}.
In the left panel, corresponding to the 150 GHz map, we identify the cluster as the signal decrement in the center of the map, characteristic of the thermal SZ effect at frequencies lower than 217 GHz.
The peak of this surface brightness decrement is detected with a $8.5\sigmaup$ significance. 
Solid black contours indicate significance levels greater than $3\sigmaup$ spaced by $1\sigmaup$.
The noise standard deviation maps used to evaluate these confidence levels are obtained by Monte Carlo realizations of noise generated from the power spectrum of residual correlated noise weighted by the associated weight map.
We note a high level of residual correlated noise, with large noise bands surrounding the SZ decrement of the cluster.
Given this noise level, we assume the contribution of the kinetic SZ effect \citep[kSZ,][]{sunyaev_ksz} to the signal to be negligible;  all further mention of the SZ effect therefore only refers to the thermal SZ effect.

We also note indications of strong contamination by point sources.
Submillimeter or radio sources indeed emit a flux that can compensate for the SZ decrement and create holes in the apparent shape of the ICM. 
This appears to be the case in the NIKA2 150 GHz map, where positive point sources lie very close to the cluster, as well as possible holes\footnotemark. \footnotetext{The holes are positive signal that compensates for the negative SZ, and they are therefore bumps.}
This is particularly concerning in the case of this cluster because of its small spatial extension: the area with an S/N greater than $3\sigmaup$ is $\simm 1 \,\mathrm{arcmin}^2$, corresponding to $ \text{about ten}$ times the solid angle of the NIKA2 instrumental beam at 150 GHz.
A single point source near the cluster can therefore have a strong effect on the observed shape of the ICM.

In the right panel, corresponding to the 260 GHz map, no SZ is detected.
The SZ signal, which is expected to be slightly positive in this frequency channel, is not detected given the noise level.
For this cluster, the peak amplitude of the SZ signal of the cluster is expected to be $\sim 0.2 \;\mathrm{mJy/beam}$, which is $ \text{about four}$ times smaller than the instrumental noise at the center of the map.
However, this map allows us to clearly identify point sources very near the center of our cluster, indicating that a careful evaluation of the contamination by these sources must be performed before we can retrieve accurate information on the ICM.

\subsection{X-ray observations of ACT-CL J0215.4+00309}\label{subsec:xmm}

\begin{figure*}[!thp]
    \centering
    \includegraphics[height=7cm, trim={0cm, 0cm, 1cm, 1cm}, clip]{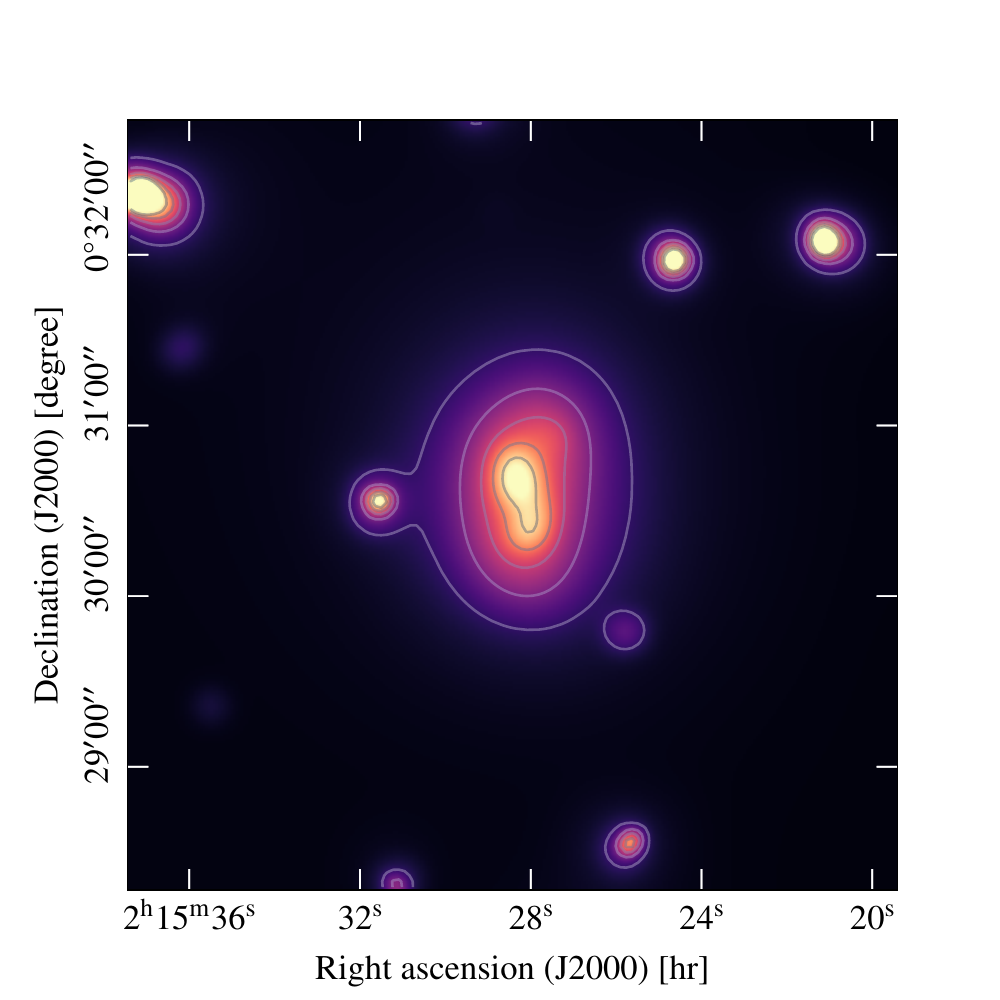}
    \hspace{20pt}
    \includegraphics[height=7cm, trim={0cm, 0cm, 1cm, 1cm}, clip]{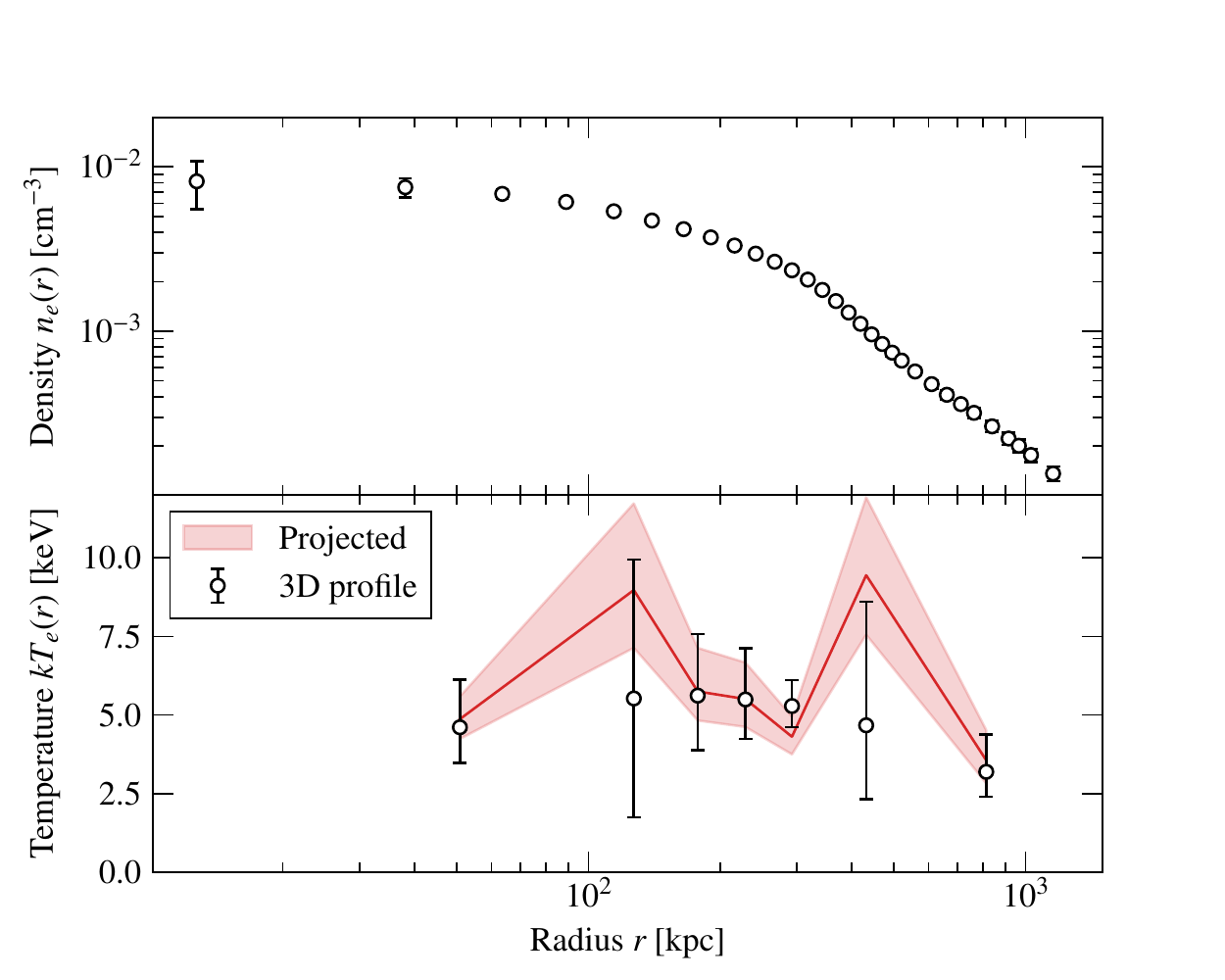}
    \caption[XMM]{%
        \textbf{Left.} Wavelet-smoothed XMM-\textit{Newton} photon-count image of the cluster.
        The region is the same as for the NIKA2 maps in Fig.~\ref{fig:nk2maps}.
        The isocontours are shown in grey.
        \textbf{Right.} Electron density (\textit{top}) and temperature (\textit{bottom}) profiles extracted from XMM observations of the cluster.
        The profiles are extracted at the SZ surface brightness peak; see \S\ref{subsec:xmm}.
        Error bars are $1\sigmaup$.
        For the temperature profile, the red line shows the projected temperature, and the white points represent the reconstructed 3D profile; see \citet{pratt_gas_2010} for a detailed explanation of the difference.
    }
    \label{fig:xmm}
\end{figure*}

ACT-CL~J0215.4+0030 was observed by XMM-Newton for a total observation time of 37 ks. 
Data (OBSID: 0762290501) were retrieved from the archive, and the standard procedures \citep[see \eg][]{bartalucci_resolving_2017} were followed to produce cleaned and calibrated event files: we applied the vignetting correction, produced background files, detected and excluded point sources, and subtracted the background. 
The observation time after cleaning was 35/29 ks (MOS/pn).

The wavelet-smoothed image of the cluster shown in Fig.~\ref{fig:xmm} indicates that the object may be somewhat disturbed. 
Deprojected PSF-corrected gas density and temperature profiles were produced from the X-ray data as described in\ \citet{pratt_gas_2010}, for example.
In view of the likely perturbed nature of the object, we produced the density and temperature profiles both for the X-ray and the SZ peak, although the resulting profiles are not markedly different. 
The density and temperature profiles, centered on the SZ peak, are shown in Fig.~\ref{fig:xmm}.
The projected temperature profile, extracted directly from 2D spectra, is also shown for reference, but only the deprojected profile was used in this paper.
The hydrostatic total mass profile, obtained under the assumption of spherical symmetry, was derived using the Monte Carlo procedure described in detail in 
\citet{democles_testing_2010} and \citet{bartalucci_resolving_2018}.
This profile is shown in Fig.~\ref{fig:profiles}.

\section{Contamination by point sources}\label{sec:ps}

As discussed in \S\ref{subsec:nk2data}, a precise estimation of the contamination by point sources is crucial in order to be able to retrieve accurate properties of the cluster through the SZ effect.
This is because the flux of point sources can partially or fully compensate for the SZ decrement and disturb the apparent shape of the cluster.
In the case of ACT-CL~J0215.4+0030, this effect is even more important because the cluster is faint (the peak surface brightness is lower than $1\;\mathrm{mJy/beam}$ at 150 GHz) and has a small spatial extension compared to the NIKA2 camera resolution.
In this section, we investigate the origin of the contamination and detail the procedure we used to estimate the amplitude of this contamination in the NIKA2 150 GHz map.

\subsection{Identification of point sources}\label{subsec:ps:id}
The SZ decrement can be affected by positive point sources that are either foreground or background ones, or are sources that belong to the cluster.
In any case, they can be dusty submillimeter galaxies (SMGs), with stronger emission at
frequencies higher than the those covered by the NIKA2 bandpasses\footnotemark, or radio sources that mostly emit at lower frequencies.
\footnotetext{High-redshift SMGs may have a peak emission in the NIKA2 bandpasses.}
The former are easier to identify in NIKA2 SZ observations because their fluxes are stronger in the 260 GHz map where we expect a low contribution of the SZ signal (see \S\ref{subsubsec:nk2_maps}), while the latter often require data from external datasets to be identified, especially if they cannot be seen directly in the SZ decrement.

In the case of ACT-CL~J0215.4+0030, at least four sources are detected with a S/N greater than 3 in the NIKA2 260 GHz map (top right panel of Fig.~\ref{fig:nk2maps}).
Their fluxes in this band appear to be higher than in the 150 GHz map, promoting SMGs rather than radio sources.
Cross-matching of their positions with sources from the \textit{Herschel} Stripe-82 Survey catalog \citep[HerS,][]{viero_herschel_2014} allowed us to confirm this hypothesis, and we also identified a fifth SMG in the northeastern part of the cluster.

\begin{figure}[tp]
    \centering
    \includegraphics[width=0.9\linewidth, trim={10cm 5cm 10cm 5cm}, clip]{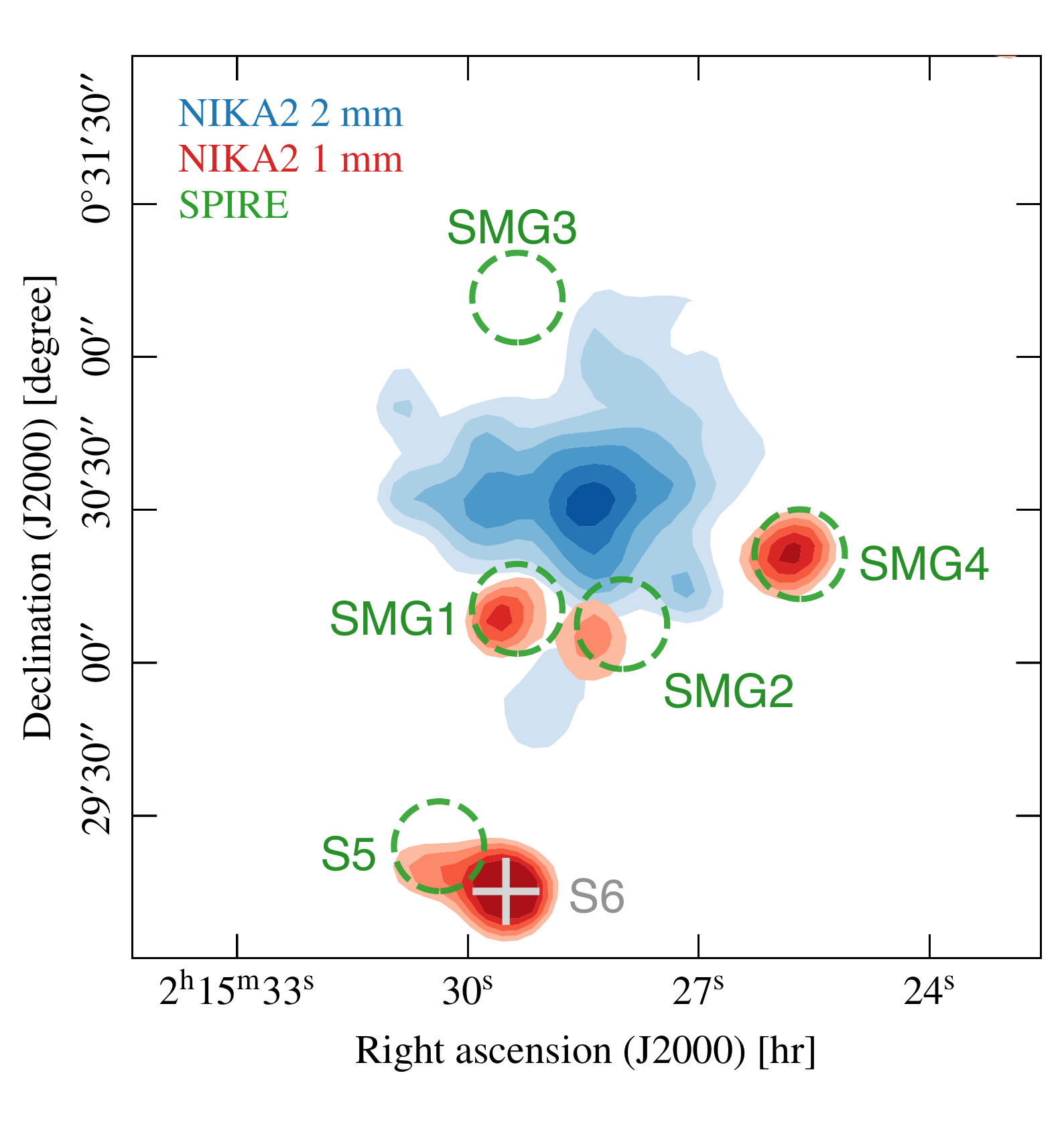}
    \caption{%
        Composite multiwavelength map of the cluster region. %
        The shaded blue regions show the NIKA2 150 GHz S/N contours, starting from 3$\sigmaup$ with increments of 1$\sigmaup$. %
        The shaded red areas represent regions with an S/N greater than 4 in the NIKA2 260 GHz map, \ie,\ the point sources. %
        The green circles are centered at the \textit{Herschel} positions of the five submillimetric point sources in our field, and their diameter is the FWHM of the SPIRE instrumental beam at 250 $\muup$m, \ie,\ 17.9''. %
        The gray gross marks the position of S6, the source identified in NIKA2 that has no counterpart in the HerS catalog (see text for details).
        }
    \label{fig:composite}
\end{figure}

Fig.~\ref{fig:composite} shows a composite multiwavelength map of the cluster region.
The green circles are centered on the positions of the five SMG identified in the HerS catalog, and the radius of each circle corresponds to the full width at half-maximum (FWHM) of the Spectral and Photometric Imaging Receiver (SPIRE) instrumental beam in the 250 $\muup$m band, which has the best resolution of all three bands.
The shaded orange areas represent the S/N of the NIKA2 260 GHz map where it is greater than 3$\sigmaup$.
 NIKA2 and SPIRE agree for the positions of the three sources in the most central part of the cluster: from left to right, SMG1, SMG2, and SMG4.
In the northeastern sector lies SMG3, the source that is identified in the HerS catalog, but has no counterpart with S/N $>3\sigmaup$ at 260 GHz in NIKA2.
In the southern region, two sources seem to be resolved in NIKA2 at 260 GHz, while only one source (hereafter S5) is identified in the HerS catalog.
A comparison of the positions of the identified sources for the two instruments shows that the source identified in HerS is the weaker of the two seen by NIKA2.
This complicates the estimation of the contamination of the SZ signal by these sources because there is no clear way to determine whether the flux of S5 in the SPIRE bands comes from a single source or from a combination of the two.
The estimation of this contamination was therefore not computed by following the same procedure as with SMG1--4 and is detailed in \S\ref{subsec:ps:more}.

\subsection{Estimation of the submillimetric contamination}\label{subsec:ps:contam}

To estimate the contamination of each source in the SZ signal, we need to know the source fluxes in the NIKA2 150 GHz band.
In this section, we describe the process we used to do so, based on the fit of the spectral energy distribution (SED) of each source.

\subsubsection{SED adjustment}\label{subsubsec:ps:contam:sed}
Each source was fit in the NIKA2 260 GHz map as a 2D Gaussian with fixed FWHM = $12.5"$, which is the reference beam pattern with which the maps are calibrated (as described in \calib).
The amplitude of the Gaussian gives us the flux of the source at the reference frequency of 260 GHz.
We then took the value of the flux of the sources in each band of the \textit{Herschel} SPIRE instrument, that is,\ 250, 350, and 500 $\muup\mathrm{m}$ (1200, 860, and 600 GHz, respectively), in the HerS catalog.
The SED of each source was adjusted using these three fluxes, combined with the measurement of their flux in the NIKA2 260 GHz band.
We used a modified blackbody spectrum model, where the SED can be written as
\begin{equation}
    F(\nu) = A_0 \left(\frac{\nu}{\nu_0}\right)^\beta B_\nu(T),
    \label{eq:sed}
\end{equation}
where $B_\nu(T)$ is the blackbody spectrum at temperature $T$, and $A_0$ is the amplitude of the SED at the frequency $\nu_0 = 500 \; \mathrm{GHz}$, $\beta$, is the spectral index of the dust, and $T$ its effective temperature, $T = T_\mathrm{dust} / (1+z)$.

The SED of each source was fit using a Monte Carlo Markov chain (MCMC) analysis \citep[and the \texttt{emcee} python package,][]{foreman-mackey_emcee:_2013} in order to obtain a complete sampling of the posterior distribution.
We emphasize the absence of submillimetric information on these sources at higher frequencies, for instance,\ from observations of the cluster region with the Photodetector Array Camera \& Spectrometer \citep[PACS,][]{poglitsch_photodetector_2010}, which were used in previous NIKA and NIKA2 studies of other clusters \citep{adam_high_2016,ruppin_first_2018}.
The degeneracy between the three SED parameters therefore affects the data strongly \citep[see, \eg,][]{desert_submillimetre_2008, magnelli_herschel_2012,
smith_isothermal_2013, berta16}, and all three of them cannot be left completely free.
We lifted this degeneracy by linearly adjusting $A_0$ in the data.
Because the spectral index and temperature of the dust are also degenerate, the prior knowledge of the spectral index $\beta$ is given by a normal distribution, $\beta \sim \mathcal{N}(2, 0.5)$, as suggested by SED measurements of large samples of galaxies \citep[see, \eg,][]{magnelli_herschel_2012}.
Flat priors are given for the temperature, that is, \ $0<T<50\;\mathrm{K}$.

At each step of the MCMC, we applied a color correction to the SPIRE fluxes by interpolating preexisting measurements\footnotemark, and to the NIKA2 260 GHz flux (see \S8.1.2 of \calib).
As an example, we show the SED of SMG1 in the left panel of Fig.~\ref{fig:ps}.
The data points used to constrain the SED (SPIRE + NIKA2 260 GHz) are shown, as well as the best-fitting SED and the 1$\sigmaup$ and 2$\sigmaup$ confidence intervals on this result.

\footnotetext{\url{http://herschel.esac.esa.int/Docs/SPIRE/html/spire_om.html}, \S5.2.7}

\subsubsection{Inference of the 150 GHz flux}\label{subsubsec:ps:contam:flux}
\begin{figure*}[tp]
    \begin{center}
    \includegraphics[width=0.48\linewidth]{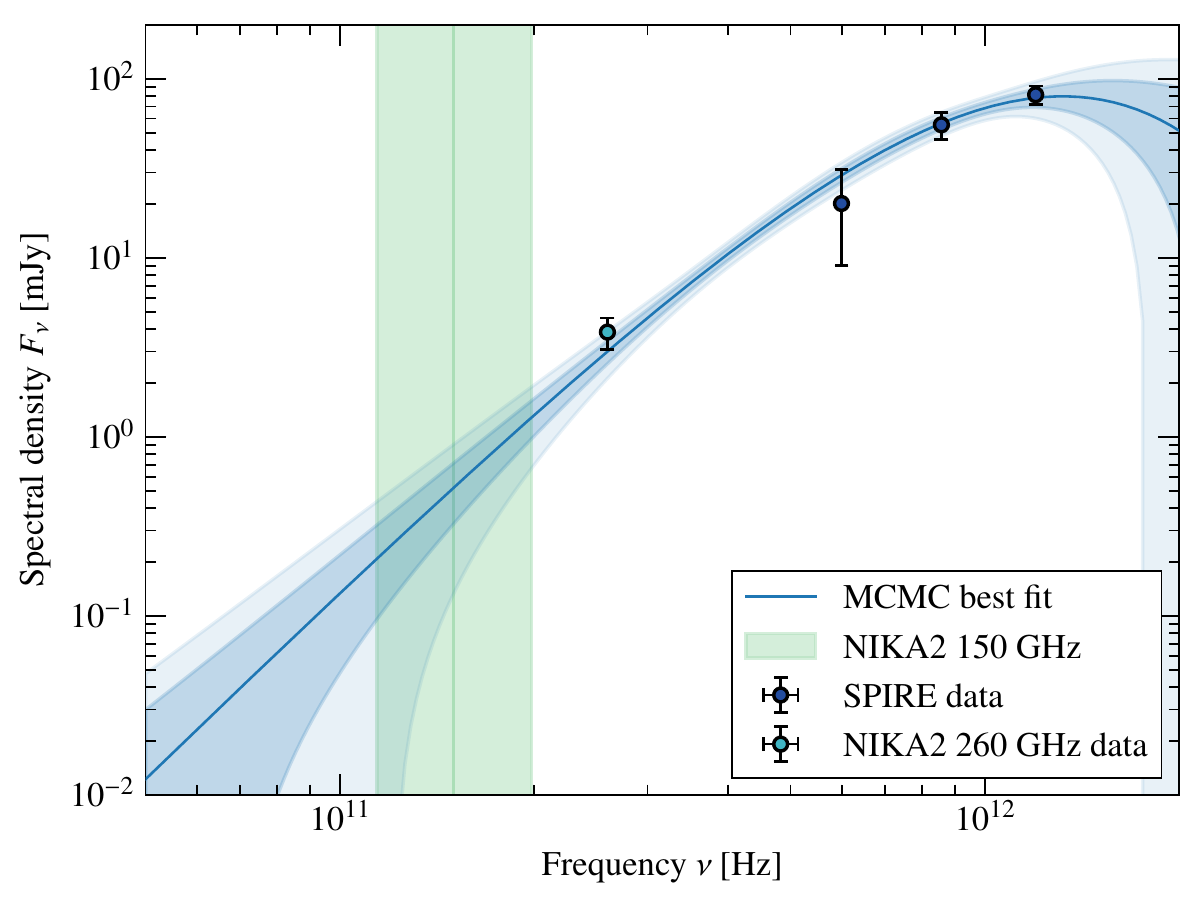}
    \includegraphics[width=0.48\linewidth]{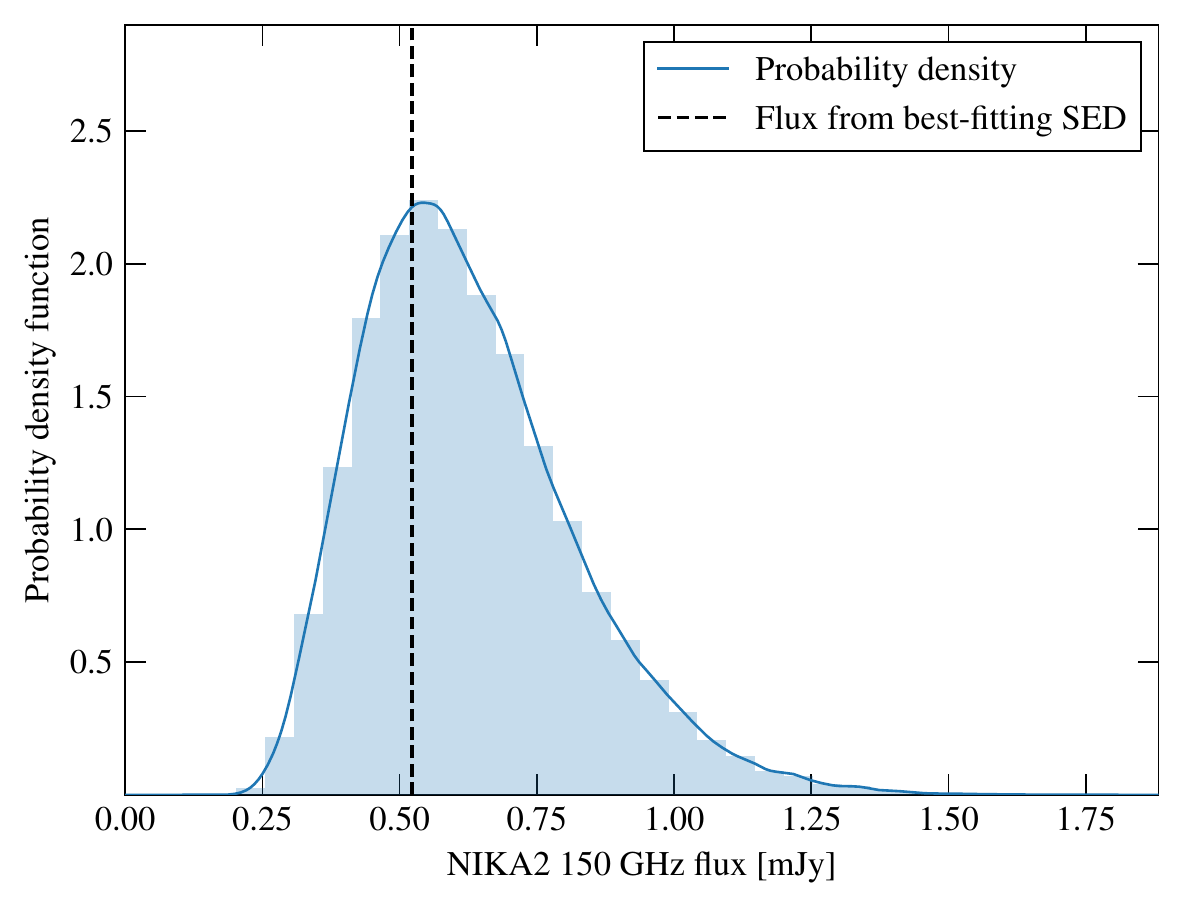}
    \caption{%
        \textbf{Left:} Spectral energy distribution of SMG1.
        The dark blue points mark the SPIRE fluxes as referenced in the HerS catalog \citep{viero_herschel_2014}.
        The light blue point shows the NIKA2 260 GHz flux measurement described in \S\ref{subsec:ps:contam}.
        The solid blue line represents the SED that best fits these four fluxes, and the blue envelopes mark the $1\sigmaup$ and $2\sigma$ confidence intervals.
        The shaded green region shows the NIKA2 150 GHz bandpass.
        \textbf{Right:} Probability distribution for the flux of SMG1 in the NIKA2 150 GHz map.
        An SED is computed for each set of parameters sampled by our MCMC and integrated in the 150 GHz bandpass to obtain a flux value at this frequency.
        The histogram in pale blue shows the distribution of the set of computed fluxes.
        Kernel density estimation is applied on the whole set of fluxes to estimate the probability distribution for the source flux in the NIKA2 150 GHz map, shown as a solid blue line.
        The flux obtained from the maximum likelihood SED is shown as a dashed black line.
        }
    \label{fig:ps}
    \end{center}
\end{figure*}

The SED adjustment outputs a sampling of the posterior probability distribution in the $(\beta, T)$ parameter space for each source.
Each sample of this distribution was then used to compute an SED through Eq. \ref{eq:sed}.
Each SED was extrapolated to compute a flux value at 150 GHz, and a color correction was applied to estimate the value of the contamination in the NIKA2 150 GHz map, as described in \calib.
Kernel density estimation was then used on the sample of fluxes to compute a probability density function (PDF) for the amplitude of the contamination of each source.
By combining this and the results of the 260 GHz fit, we have a precise knowledge of the position of each source in the NIKA2 maps, and have a PDF for its flux in the NIKA2 SZ map.
The results of this computation for SMG1 are shown in the right-hand panel of Fig.~\ref{fig:ps}.
The SED results illustrated in the left-hand panel are extrapolated at 150 GHz and color-corrected to obtain a distribution of fluxes, which is used to compute the probability distribution function for the flux of the source.

\subsubsection{Results}\label{subsec:ps:results}
We report in Table \ref{tab:pscat} the positions and fluxes of the six sources contaminating our SZ signal.
The last column reports the 150 GHz flux inferred from the extrapolation of the SED of each source in the NIKA2 150 GHz bandpass \citep[][\calib]{adam_nika2_2018}, except for the last two sources.
We see that these fluxes can be as high as 0.74 mJy, while the peak SZ flux of the cluster in the NIKA2 map is below 0.8 mJy.
This means that when this contamination is accounted for in our analysis of the SZ signal, we need to be very careful.

\begin{table*}[tp]
    \caption{Catalog of the positions and fluxes of the six submillimeter sources within a radius of 2.5' around the center of our cluster. 
    }
    \begin{center}
    \begin{tabular}{ccccccc}
        \toprule
        Source & Coordinates J2000 & 1200 GHz & 860 GHz & 600 GHz & 260 GHz & 150 GHz \\
            &  & [mJy] & [mJy] & [mJy] & [mJy] & [mJy] \\
        \midrule
        SMG1 & $02^\mathrm{h}15^\mathrm{m}29.5^\mathrm{s}$ $+00^\circ30{}^\prime07.2{}^{\prime\prime}$ & 81.55 $\pm$ 9.77 & 55.39 $\pm$ 9.63 & 20.14 $\pm$ 11.08 & 3.85 $\pm$ 0.77 & 0.62 $\pm$ 0.19 \\
        SMG2 & $02^\mathrm{h}15^\mathrm{m}28.3^\mathrm{s}$ $+00^\circ30{}^\prime03.5{}^{\prime\prime}$ & 56.57 $\pm$ 9.83 & 46.87 $\pm$ 9.83 & 23.07 $\pm$ 10.57 & 2.64 $\pm$ 0.64 & 0.38 $\pm$ 0.14 $^{(a)}$\\
        SMG3 & $02^\mathrm{h}15^\mathrm{m}29.1^\mathrm{s}$ $+00^\circ31{}^\prime10.2{}^{\prime\prime}$ & 30.19 $\pm$ 9.68 & 29.45 $\pm$ 9.69 & 13.13 $\pm$ 10.6 & 1.42 $\pm$ 0.52 & 0.21 $\pm$ 0.09 \\
        SMG4 & $02^\mathrm{h}15^\mathrm{m}25.7^\mathrm{s}$ $+00^\circ30{}^\prime21.3{}^{\prime\prime}$ & 34.6 $\pm$ 9.86 & 25.84 $\pm$ 10.11 & 16.58 $\pm$ 10.43 & 4.42 $\pm$ 0.83 & 0.74 $\pm$ 0.25 \\
        S5 & $02^\mathrm{h}15^\mathrm{m}30.0^\mathrm{s}$ $+00^\circ29{}^\prime12.6{}^{\prime\prime}$ & -- & -- & -- & -- & 0.66 $\pm$ 0.19 $^{(b)}$\\
        S6 & $02^\mathrm{h}15^\mathrm{m}29.0^\mathrm{s}$ $+00^\circ29{}^\prime14.2{}^{\prime\prime}$ & -- & -- & -- & -- & 0.61 $\pm$ 0.19 $^{(b)}$\\
        \bottomrule\\
    \end{tabular}
    \end{center}
    \vspace{-10pt}
    \footnotesize\textbf{Notes.} The 1200, 860, and 600 GHz flux densities are given by the \textit{Herschel} Stripe 82 catalog \citep{viero_herschel_2014}. %
    The coordinates and 260 GHz fluxes are obtained from the fit of the sources in the NIKA2 260 GHz map, see \S\ref{subsubsec:ps:contam:sed}.
    The 150 GHz fluxes and their error bars are the means and standard deviations of the probability distributions of the 150 GHz fluxes, see \S\ref{subsubsec:ps:contam:flux}.
    $^{(a)}$ Flux including the radio emission described in \S\ref{subsec:ps:more}.
    $^{(b)}$ Fluxes measured directly in the NIKA2 150 GHz map, as explained in \S\ref{subsec:ps:more}.
    \label{tab:pscat}
\end{table*}

\subsection{Additional considerations}\label{subsec:ps:more}
\paragraph{\textit{Radio contamination.}} One of the sources (SMG2) also emits in the radio wavelengths.
It is detected in the latest FIRST catalog \citep{helfand_last_2015} with a $2.22 \pm 0.10 \;\mathrm{mJy}$ flux density at 1.4 GHz.
The lack of coverage of the cluster region in other radio frequencies prevents us from fitting the SED radio part of this source.
We modeled it as a power law $F(\nu) = F(\nu_0)\times(\nu/\nu_0)^\alpha$, with a spectral index $\alpha=-0.7 \pm 0.2$, which was found to be a good description of most radio galaxies \citep{condon_cosmological_1984}.
The extrapolation of this power law into the NIKA2 bandpass gives an estimate of the flux of the galaxy due to radio emission, $F_\mathrm{radio}(\mathrm{150\; GHz}) = 83 ^{+127}_{-50} \,\mathrm{\muup Jy}$.
This contribution was added to the contribution computed from the submillimeter prediction to infer the total flux of the source at 150 GHz, although it only accounts for a small fraction of the total.

\paragraph{\textit{Double southern source.}} As discussed in Sect. \ref{subsec:ps:id}, the complex structure in the southern part of the map appears to be two sources in NIKA2, but only one is identified in the HerS catalog.
We label these sources S5 and S6 in Table \ref{tab:pscat}.
Although the position-matching appears to indicate that the source in the HerS catalog is the fainter of the two seen by NIKA2, we cannot be sure that this is the case.
We directly measured the fluxes of these two sources in the NIKA2 150 GHz map because they lie far enough away from the cluster that no SZ signal is expected.
The recovered fluxes of the two sources in the NIKA2 150 GHz map, as well as their position in this map, are reported in Table \ref{tab:pscat}.
They were subtracted from the 150 GHz map in order to avoid a bias in the fit of the SZ signal (see \S\ref{sec:panco}).

\section{ICM reconstruction in the MCMC analysis}\label{sec:panco}

In this section, we present the procedure we used to derive thermodynamical properties of the ICM from the NIKA2 map.
We used the NIKA2 SZ pipeline to do this. The procedure is described by \citet{ruppin_first_2018}.
We adapted this pipeline in order to better account for the contamination by point sources described in \S\ref{sec:ps}.

\subsection{Pressure profile}\label{subsec:pressprof}
As discussed earlier, the amplitude of the SZ effect, and therefore the surface brightness of a cluster observed in the millimeter domain,  is proportional to the electron pressure integrated along the line of sight (Eq. \ref{eq:szy}).
Millimeter maps of a cluster can therefore be used to constrain the pressure distribution in the ICM.
When spherical symmetry is assumed, this distribution can be described by a radial pressure profile of the ICM.
This pressure profile was integrated along the line of sight to compute a map of the Compton parameter of the cluster.
We modeled the pressure distribution with a generalized Navarro-Frenk-White profile \citep[gNFW,][]{nagai_effects_2007},~
\begin{equation}
    P_e(r) = P_0 \left(\frac{r}{r_p}\right)^{-c} \left[1 + \left(\frac{r}{r_p}\right)^a\right]^{(c-b)/a},
    \label{eq:gnfw}
\end{equation}
where $b$ and $c$ are the external and internal slope of the profile, respectively, $r_p$ is the characteristic radius of the transition between the two regimes, $a$ is the steepness of the transition, and $P_0$ is a normalization constant.

\subsection{MCMC analysis}\label{subsec:mcmcpanco}
We used MCMC sampling to fit the pressure profile of ACT-CL~J0215.4+0030.
At each step of the MCMC, a pressure profile was computed using Eq. \ref{eq:gnfw} and integrated along the line of sight (Eq. \ref{eq:szy}) to derive a Compton parameter $y$ profile.
This profile was projected on the same pixelated grids as the NIKA2 map to compute a Compton parameter map that we compared with the data.
The resulting $y$-map was convolved with the NIKA2 beam, that is,\ a 2D Gaussian function with ${\rm FWHM=18.5"}$.
The $y$-map was centered on the coordinates of the peak in SZ surface brightness in the NIKA2 150 GHz map.

The $y$-map was then converted into surface brightness units using a conversion coefficient $C_\mathrm{conv}$.
This coefficient depends on the NIKA2 150 GHz bandpass, but also on the atmospheric opacity and the shape of the SZ spectrum.
Therefore, the coefficient was adjusted in our fit with a Gaussian prior with a standard deviation of 10\% in order to take the uncertainty on the absolute calibration of the NIKA2 data into account.

In order to consider the point source contamination, we fit each source simultaneously with the pressure of the ICM rather than using the fluxes obtained by SED extrapolation and subtracting the sources in our NIKA2 150 GHz map.
To do so, 2D Gaussian functions corresponding to the NIKA2 150 GHz beam pattern $\mathcal{B}_{150}$ were added to the NIKA2 model.
Their amplitudes were left as free parameters.
A prior on the flux of each source is given by the probability density obtained from the SED extrapolation (see \S\ref{subsec:ps:contam}).
This has the advantage of naturally taking the uncertainty on the extrapolated source fluxes into account.
Sources S5 and 6, for which we were unable to fit and extrapolate am SED, were directly subtracted from the data using their fluxes and positions measured in the NIKA2 150 GHz map (see \S\ref{subsec:ps:more}).
Because they are located far from the cluster (outside $R_{500}$), this reduces the number of parameters of our MCMC analysis without altering the quality of the recovered profiles.
The resulting surface brightness map was then convolved with the transfer function to account for filtering effects due to data processing.
We also ran the joint fit with flat priors on the point source fluxes in order to ensure that our results were independent of priors and that the fluxes recovered by the joint fit without prior were compatible with those extrapolated from the SED adjustment of each source (see \S\ref{ap:psflux}).

Our ability to constrain the pressure profile in the outskirts of the cluster, represented in our model by the external slope of the pressure profile $b$, is limited by the angular coverage of NIKA2.
Constraints on the large-scale emission can be obtained by using the integrated SZ signal within an aperture of radius $R_{500}$,
\begin{equation}
    \mathcal{D}_\mathrm{A}^2 \; Y_{500} = 4\pi \frac{\sigma_\textsc{t}}{m_e c^2}\int_0^{R_{500}} r^2 P_e(r) \, \dd r,
    \label{eq:y500}
\end{equation}
where $\mathcal{D}_\mathrm{A}$ is the angular diameter distance at the cluster redshift.
The integrated Compton parameter was measured in the ACT survey, $\mathcal{D}_\mathrm{A}^2 \, Y_{500}^\mathrm{ACT} = (4.07 \pm 1.13) \times 10^{-5} \;\mathrm{Mpc}^2$ \citep[][hereafter \act]{hasselfield_atacama_2013}.
We took this into account in our likelihood.
For each set of parameters sampled in our MCMC, the value of $Y_{500}$ was computed and compared to that of the ACT survey.

To summarize, at each iteration of the MCMC, the model map was computed from the following set of parameters $\vartheta$:
\begin{itemize}
    \setlength\itemsep{5pt}
    \item $P_0$, $r_p$, $a$, $b$, and $c$: the parameters describing the gNFW pressure profile of the ICM (Eq. \ref{eq:gnfw}), with Gaussian priors around the \upp\ universal pressure profile values,
    \item $F_{150}^i$, $i = 1 \dots 4$: the fluxes of every point source near our cluster, with priors given by the extrapolation of the SED of sources in the NIKA2 bandpass (\S\ref{subsec:ps:contam}),
    \item $C_\mathrm{conv},\,Z$: the $y$-to-Jy/beam conversion coefficient and the zero-level of the map.
\end{itemize}
The probability that these parameters describe our data was obtained by combining our prior knowledge of the data with the likelihood that compared model $\mathcal{M}$ with data $\mathcal{D}$,
\begin{align}
    \nonumber\mathrm{log}\,\mathcal{L}(\vartheta) =& -\frac{1}{2}\sum_{i=1}^{n_\text{pixels}^\text{NIKA2}} \bigg[\Big(\mathcal{M}(\vartheta) - \mathcal{D} \Big)^T C^{-1} \Big(\mathcal{M}(\vartheta) - \mathcal{D} \Big)\bigg]_i \\
    & - \frac{1}{2} \left(\frac{Y_{500}^\mathrm{ACT} - Y_{500}(\vartheta)}{\Delta Y_{500}^\mathrm{ACT}}\right)^2,
    \label{eq:likelihood}
\end{align}
where $C$ the noise covariance matrix, which is evaluated by computing the covariance of
Monte Carlo noise realizations generated from the correlated noise power spectra
\citep[see Appendix~\ref{ap:noisetf}, as well as previous NIKA(2) papers, \eg,][]{adam_high_2016,ruppin_first_2018}.

The MCMC analysis was performed using the \texttt{emcee} python package \citep{foreman-mackey_emcee:_2013}.
The convergence of the chains was monitored using the $\hat{R}$ test of \citet{gelmanrubin}.
The sampling was performed using 30 walkers.
A burn-in time of 400 iterations was removed from the chains, leaving $4 \times 10^5$ points in the final posterior distribution.

\subsection{Consistency of point source flux estimations}\label{ap:psflux} 
\begin{figure}[t]
    \begin{center}
	\includegraphics[width=\linewidth, trim={.35cm .5cm .35cm 1.4cm}, clip]{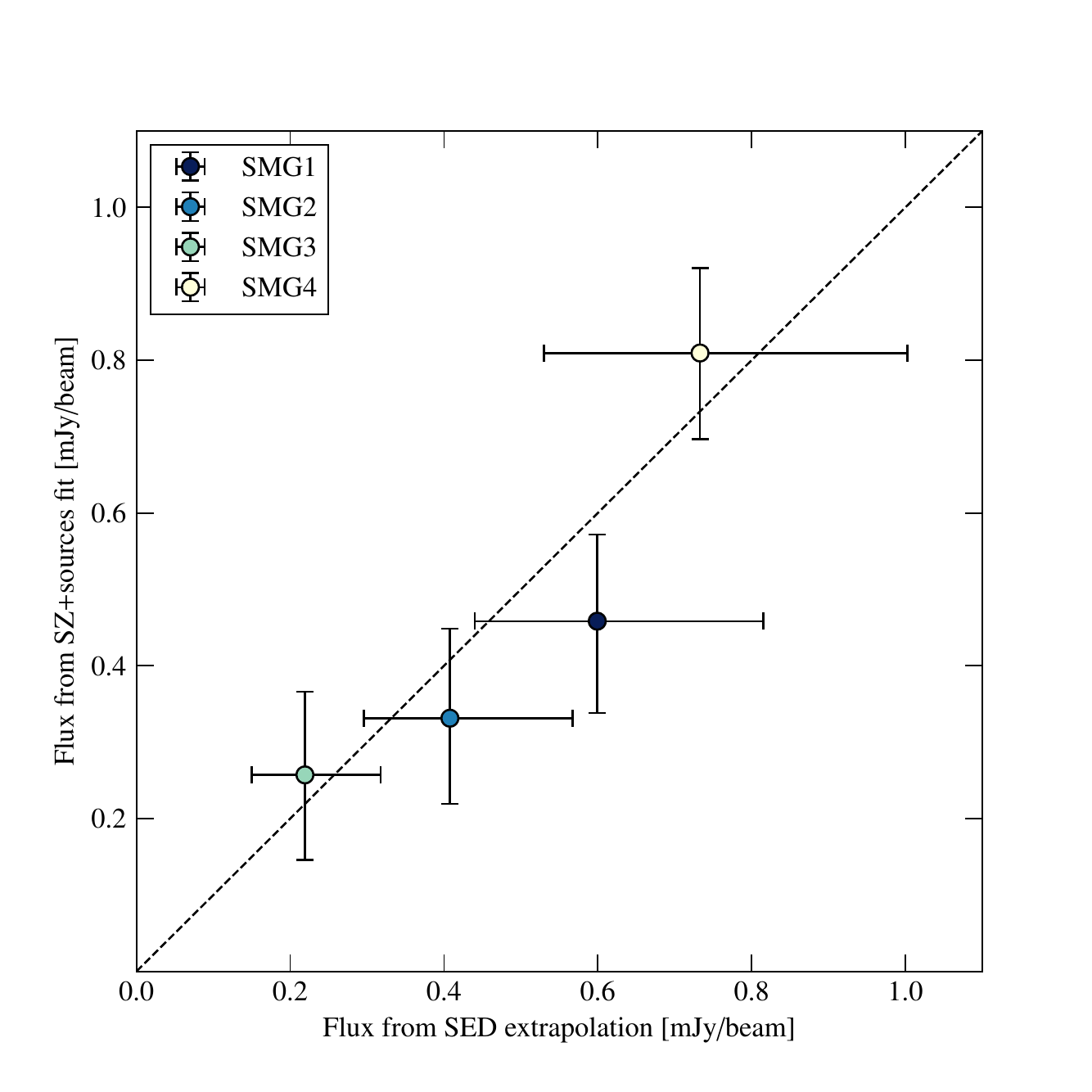}
    \caption{%
                Comparison of the fluxes of point sources estimated by SED extrapolation ($x$ -axis) and by point sources + ICM reconstruction analysis ($y$ -axis).
        The error bar edges represent the 16th ant 84th percentile of the distributions.
        The two flux estimators are compatible within the error bars for each point source.
    }
    \label{fig:pscomp}
    \end{center}
\end{figure}

The parameters evaluated by our MCMC reconstruction of the ICM include the fluxes of the four point sources near our cluster.
We compared these measurements with those obtained by the SED extrapolation described in \S\ref{sec:ps}.
The results are found to be compatible, but because we used the SED extrapolation results as priors for the ICM reconstruction, the two measurements are not strictly independent.
In order to be able to run the comparison, we repeated our MCMC reconstruction of the ICM with uninformative priors on the flux of each source.
The posterior distribution becomes independent of the SED extrapolation measurement, and the two estimates of point source fluxes can be compared.
The results are shown in Fig.~\ref{fig:pscomp}.
The results of both estimations of the point sources fluxes are compatible within $1\sigmaup$.
This gives confidence in our estimation of the fluxes, and therefore of the SZ signal in the contaminated regions of the map.
Moreover, although this shows that no information is gained from using the SED extrapolation as priors for the joint ICM+point source fit, the priors greatly reduce the time needed for the MCMC to converge, reducing the available sampling volume in the parameter space.

\section{Results}\label{sec:results}
\begin{figure*}[!tp]
    \begin{center}
    \includegraphics[width=.95\linewidth]{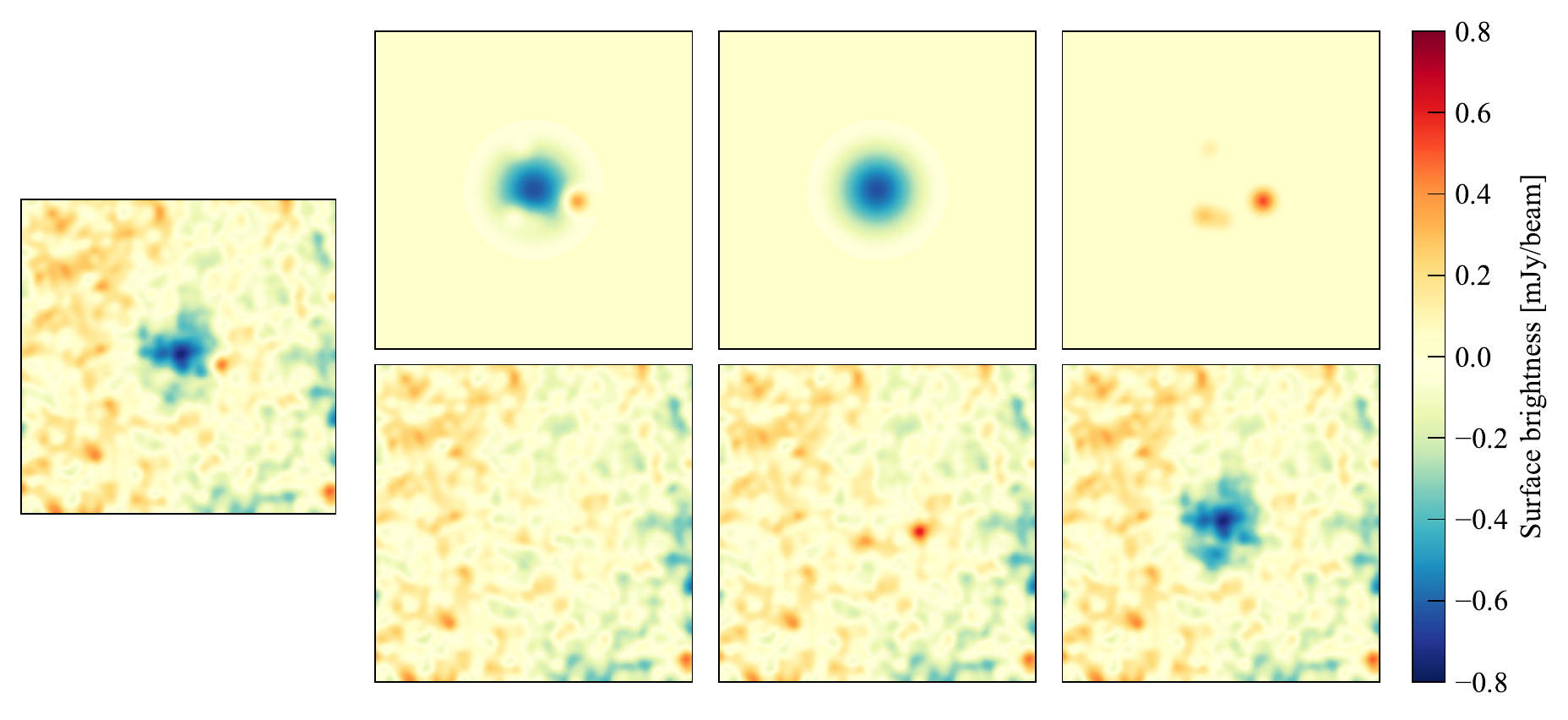}
    \caption{%
             Results of the fit of the NIKA2 SZ map of ACT-CL J0215.4+0030.
             The far left panel is the NIKA2 150 GHz map.
             The best-fitting model and residuals are represented in the top and bottom panels of the center left column, respectively.
             The center right and far right columns show the best-fit model and residuals considering only the SZ signal and only the point sources.
             The NIKA2 maps are given in the same 4.5'$\times$4.5' area with the same color scale and smoothed with a 10'' Gaussian kernel for display purposes.
             The two southern sources have been subtracted from the NIKA2 map (see \S\ref{subsec:mcmcpanco}).
        }
    \label{fig:dmr}
    \end{center}
\end{figure*}

\begin{figure}[thp]
    \begin{center}
    \includegraphics[width=\linewidth, trim={0cm .2cm 0cm 1cm}, clip]{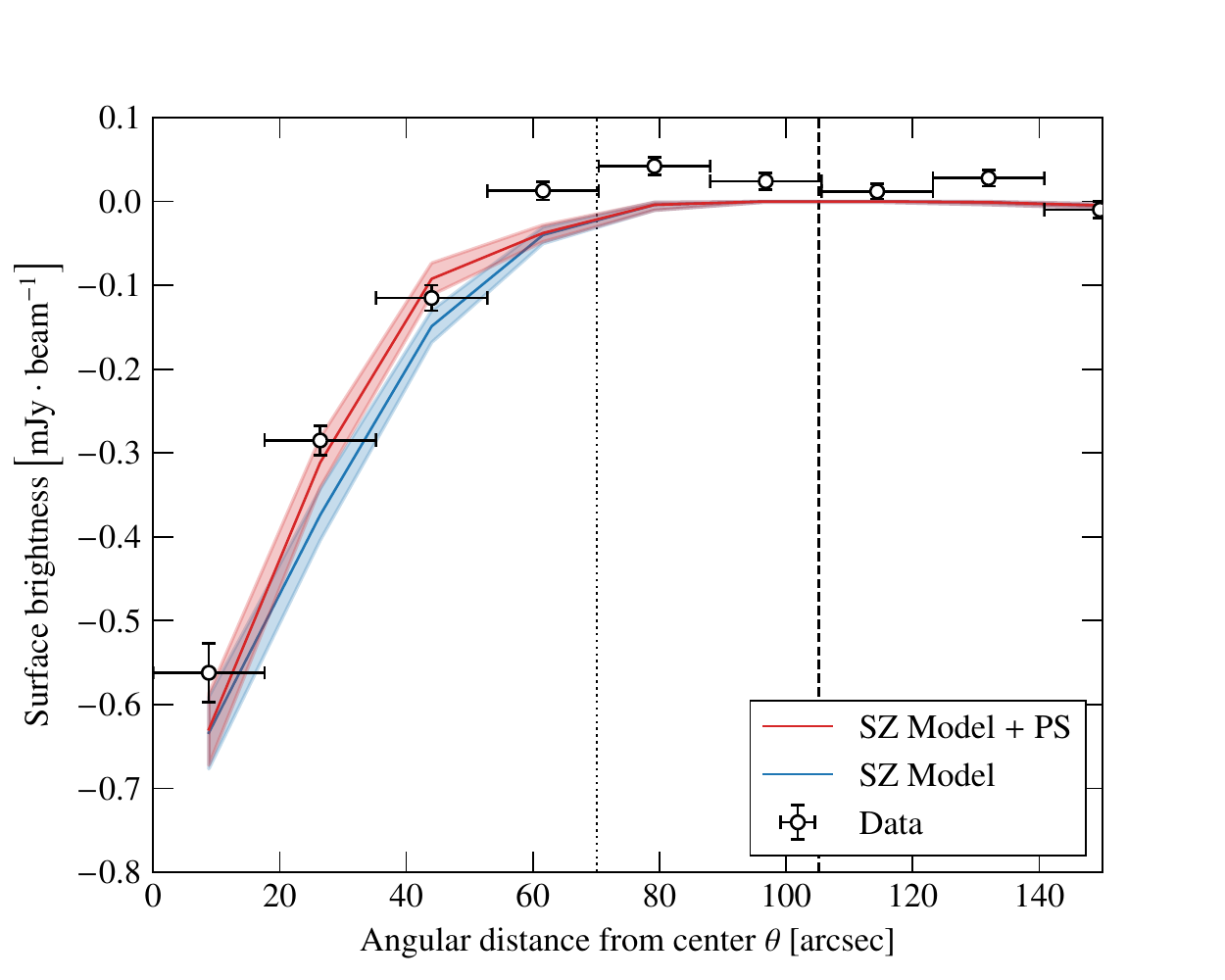}
    \caption{%
        Projected radial profiles evaluated in concentric annuli centered on the SZ peak, with widths equal to the NIKA2 instrumental FWHM at 150 GHz.
        The white points show the radial profile of the NIKA2 SZ signal with $1\sigmaup$ errors.
        The red curve corresponds to the radial profile of our full SZ + point sources model, and the blue curve is the profile of the SZ signal alone.
        In each case, the envelopes are $1\sigmaup$ and show the dispersion in the posterior distribution sampled by our MCMC.
        The dotted vertical line shows the limit of the region in which the S/N on the surface brightness profile is $3\sigmaup$.
        The dashed black line shows the angle subtended by $R_{500}$ , as reported in Table~\ref{tab:integ}.
        We emphasize that this is only a comparison between the angular radial profile and the best fit on the 2D map, and that the fit is indeed performed in 2D; see \S\ref{subsec:mcmcpanco}.
        }
    \label{fig:yprof}
    \end{center}
\end{figure}

The MCMC procedure described in the previous section was used to determine the parameters that best fit the NIKA2 150 GHz map as the sum of the SZ signal for an ICM described by a gNFW pressure profile and of contamination by four point sources, convolved by the NIKA2 instrumental response and by the data processing transfer function.
The 2D model and residuals are shown in Fig.~\ref{fig:dmr}.
Because no high-S/N structures are seen in the residuals map, except  for the western structure at the edge of the map, well outside the cluster, we can say that there is no evidence for substructures or departure from sphericity in the NIKA2 data.
In addition, Fig.~\ref{fig:yprof} shows the surface brightness profiles of the NIKA2 map and of the model.
The two profiles are not significantly offset, further indicating that the gNFW + point sources model describes our data well.
We underline that this figure is only an illustration of the compatibility of the data and model in 1D, and that the fit was performed on the NIKA2 map, not on the radial profile.

\subsection{Thermodynamic profiles}\label{subsec:profiles} 

\begin{figure*}[!tp]
    \centering
    \includegraphics[page=1, width=.48\linewidth, trim={.3cm .1cm .9cm .5cm}, clip]{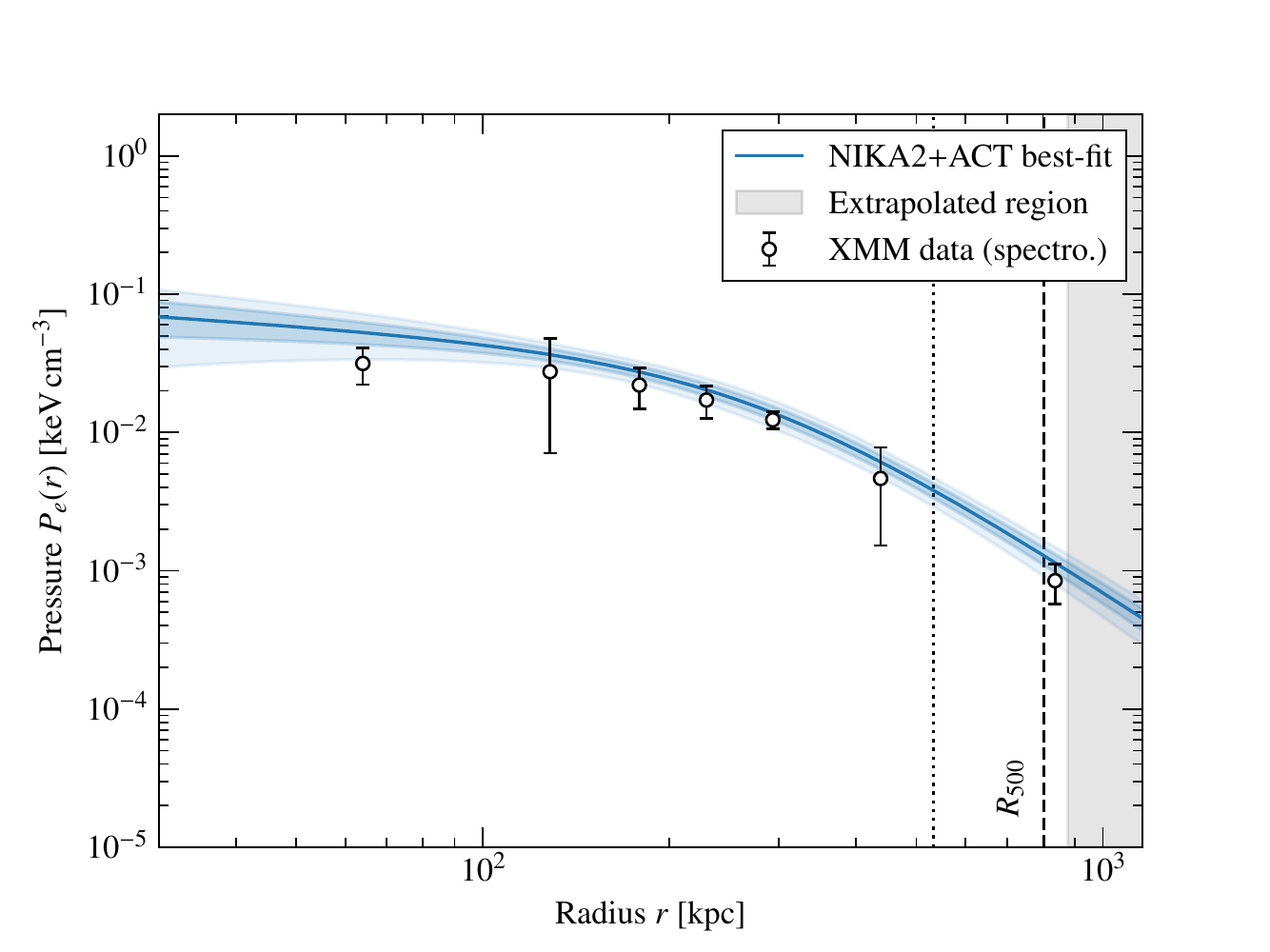}\hfill
    \includegraphics[page=2, width=.48\linewidth, trim={.3cm .1cm .9cm .5cm}, clip]{Figures/ICM_thermodynamics.pdf}
    \includegraphics[page=3, width=.48\linewidth, trim={.3cm .1cm .9cm .5cm}, clip]{Figures/ICM_thermodynamics.pdf}\hfill
    \includegraphics[page=4, width=.48\linewidth, trim={.3cm .1cm .9cm .5cm}, clip]{Figures/ICM_thermodynamics.pdf}
    \caption{%
             Radial profiles of the ICM pressure (\textit{top left}), temperature (\textit{top right}), entropy (\textit{bottom left}), and hydrostatic mass (\textit{bottom right}) of ACT-CL J0215.4+0030.
             For the pressure profile, the blue line shows the maximum likelihood profile from the NIKA2+ACT analysis.
             For the other profiles, it shows the profile obtained by combining the maximum likelihood pressure profile from NIKA2+ACT with the density from XMM-\textit{Newton}.
             The blue envelopes show the $1\sigmaup$ and $2\sigmaup$ confidence intervals on these profiles.
             The white data points show the profiles obtained by combining XMM-\textit{Newton} data with and without spectroscopy with $1\sigmaup$ errors.
             The dotted black line shows the limit of the region in which the S/N on the NIKA2 surface brightness profile is greater than 3.
             The shaded gray area shows the region in which the profiles are unconstrained by SZ data, \ie,\ beyond the measurement of $R_{500}^\mathrm{ACT}$. 
        }
    \label{fig:profiles}
\end{figure*}

The pressure profile obtained from our MCMC adjustment is presented in the top left panel of Fig.~\ref{fig:profiles}.
The blue line marks the best-fitting gNFW profile.
The error envelopes were obtained by computing a pressure profile for each set of sampled parameters after a burn-in cutoff and computing the dispersion of these profiles.

We compared this pressure profile with that inferred from the combination of X-ray density and temperature profile (Fig.~\ref{fig:xmm}).
The comparison between these two profiles is interesting because they represent two independent measurements of the pressure distribution in the ICM in the specific case of a distant cluster.
The X-ray only data are superimposed in white in the top left panel of Fig.~\ref{fig:profiles}.
The two measurements agree within the error bars.
The pressure recovered by NIKA2 appears to be higher than the X-ray only pressure in the central region of the ICM, but this effect is not significant given the error bars on both profiles.

Other thermodynamic quantities can be derived by combining this pressure profile with the density profile obtained from X-ray data without spectroscopic information.
We can compute the temperature $T_e(r)$, entropy $K_e(r),$ and hydrostatic mass $M_\mathrm{HSE}(<r)$ profiles through 
\begin{equation}
    k_\textsc{b} T_e(r) = \frac{P_e(r)}{n_e(r)}, \quad K_e(r) = P_e(r)\;n_e^{-5/3}(r),
\end{equation}
and through the equation of hydrostatic equilibrium,
\begin{equation}
    M_\mathrm{HSE}(<r) = -\frac{1}{\mu m_p G}\frac{r^2}{n_e(r)}\frac{\dd P_e}{\dd r},
    \label{eq:mhse}
\end{equation}
where $m_p$ is the proton mass, $\mu$ is the mean molecular weight of the gas, and $G$ is the gravitational constant.

We used the pressure profile recovered from the NIKA2 data in combination with the density profile from XMM-\textit{Newton} observations.
The density profile was interpolated by a power law in order to compute the value of the density at any given radius.
Fig.~\ref{fig:profiles} shows the constraints placed on the temperature, entropy, and hydrostatic mass profiles (in blue) that can be compared with the same profiles inferred using X-ray only data with spectroscopy (in white). 
As for the pressure, the profiles agree very well, illustrating the complementarity of high-resolution X-ray and SZ data.

The agreement between the thermodynamic profiles recovered with NIKA2+XMM and XMM-only observations is a great assessment of the possibility of retrieving quality information on the ICM with NIKA2.
While the cluster was observed for comparable times with the two instruments, this compatibility shows that measurements of great quality can be performed by combining high-resolution SZ and X-ray data, which minimizes our reliance on time-expensive X-ray spectroscopy measurements.

Combined high angular resolution SZ and X-ray data also enable a better resolution for the temperature profile than can be obtained through X-ray spectroscopy alone.
Fig.~\ref{fig:profiles} shows that extrapolating the inner slope of the XMM-only temperature profile indicates a significantly lower core temperature value than is obtained with the combination of NIKA2+XMM.
The error bars on the temperature profile are also smaller on average for the NIKA2+XMM profile.
Consequently, the entropy and mass profiles obtained from the SZ+X-ray combination display smaller error bars, leading to a more precise measurement of the thermodynamical properties of the ICM.

\begin{figure}[tp]
    \begin{center}
    \includegraphics[width=\linewidth, trim={.35cm .2cm .35cm 1cm}, clip]{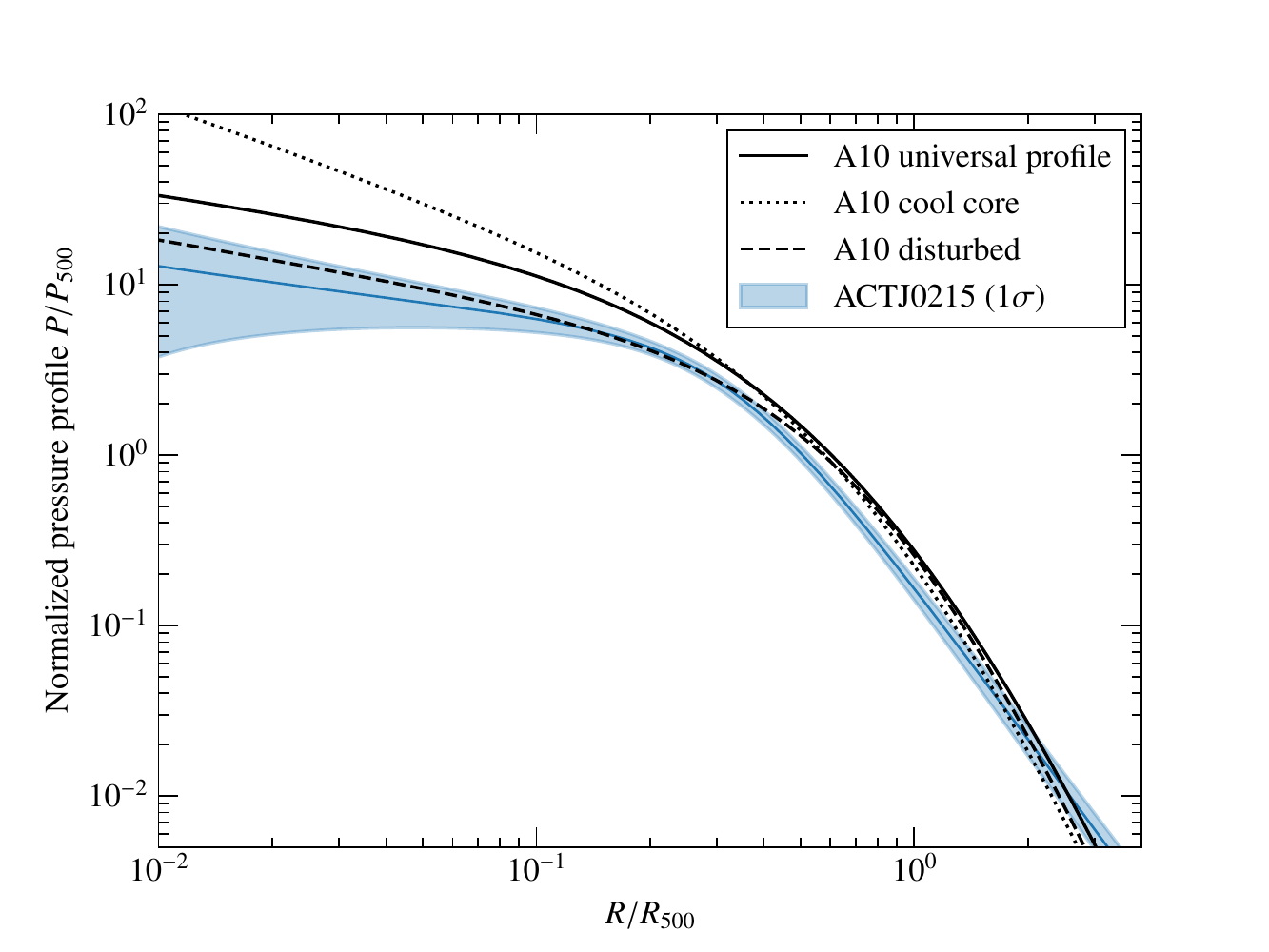}
    \caption{%
        Comparison of the deprojected pressure profile obtained from the NIKA2 data with the universal pressure profiles from \upp.
        ACT-CL~J0215.4+0030 exhibits a pressure profile similar to those identified as disturbed in \upp\ (dashed black line).
        }
    \label{fig:press_upp}
    \end{center}
\end{figure}

The thermodynamic profiles give us access to information about the dynamical state of the ICM.
The flatness of the pressure profile in the inner part of the cluster indicates a disturbed core.
The central value of the entropy profile ($K_{e,\mathrm{core}} \gtrsim 200\;\mathrm{keV\cdot cm^2}$) also indicates a noncool core cluster \citep[see, \eg,][]{hudson_what_2010}.
We compare the deprojected pressure profile with the profiles of \upp\ in Figure \ref{fig:press_upp}.
The pressure profile is poorly described by the universal pressure profile of \upp, especially in the inner parts of the cluster.
However, the ICM of ACT-CL~J0215.4+0030 appears to be well described by the mean pressure profile of disturbed clusters in the REXCESS sample (\upp).
This further indicates disturbed core dynamics of the ICM.
Similar conclusions have been drawn by \citet{ricci_xxl_2020}, who used NIKA2 to image another distant and low-mass cluster.

\subsection{Integrated quantities}\label{subsec:integ} 
For low angular resolution surveys, the thermodynamic profiles of the ICM of high-redshift clusters cannot be resolved.
Therefore the scaling relation used to compute the mass of a cluster links its integrated SZ signal $Y$ to the mass enclosed within a given radius.
We chose the radius $R_{500}$, corresponding to the radius enclosing an average density 500 times greater than $\rho_c(z)$, the critical density of the Universe at the redshift of the cluster.
Its value can be computed as the radius for which the overdensity contrast $\delta_c$ is equal to 500, that is,\ by solving
\begin{equation}
    \delta_c(R_{500}) = \frac{M_\mathrm{HSE}(<R_{500})}{\rho_c(z) \times \frac{4}{3} \pi R_{500}^3} = 500.
\end{equation}
The recovered value is
\begin{equation}
    R_{500} = 810.1 \pm 41.9 \, \mathrm{kpc},
\end{equation}
in agreement with the ACT measurement of \act.
The integrated SZ signal inside this radius, $Y_{500}$ via Eq.~\ref{eq:y500}, 
\begin{equation}
    \mathcal{D}_\mathrm{A}^2 \; Y_{500} = (3.76 \pm 0.39)\times 10^{-5} \,\mathrm{Mpc^2}.
\end{equation}
The hydrostatic mass of the ICM enclosed within a radius of $R_{500}$ is obtained through Eq.~\ref{eq:mhse},
\begin{equation}
    M^\mathrm{HSE}_{500} = M_\mathrm{HSE}(<R_{500}) = (3.79 \pm 0.58) \times 10^{14} \,\mathrm{M}_\odot.
\end{equation}

These results are shown in Table \ref{tab:integ}.
For comparison purposes, we also show two measurements of radius and mass for this cluster:
from the ACT galaxy clusters catalog \act, where $R_{500}$ and $M_{500}$ are computed from the match-filtering detection of clusters in the ACT maps assuming an \upp\ pressure profile, and from the stand-alone analysis of XMM-\textit{Newton} data.
The uncertainty on our measurement of $Y_{500}$ is reduced by a factor of 3 compared to the ACT measurement, indicating that a significant gain in precision is obtained from the high-resolution observations.
The last line of Table~\ref{tab:integ} reports the measurement of the hydrostatic mass of the cluster obtained by combining the SZ pressure profile and the X-ray density (Eq.~\ref{eq:mhse}).
This mass is compatible with the masses obtained from the $Y_{500}-M_{500}$ scaling relation that was used to build the ACT catalog of \act.
The error bar on the NIKA2+XMM mass measurement is smaller than that of the ACT catalog, even though the latter is underestimated because it does not include the uncertainty on the scaling relation regression (see Table~8 of \act).
Our mass measurement is also compatible with the mass obtained from the stand-alone X-ray analysis, although the latter is slightly lower.

\begin{table}[tp]
    \caption{Characteristic values of ACT-CL J0215.4+0030 from the analysis of ACT, XMM-\textit{Newton,} and NIKA2+XMM-\textit{Newton} data.}
    \begin{center}
    \small
    \begin{tabular}{c c c c}
        \toprule
        &  ACT  &  XMM-\textit{Newton}  &  NIKA2+XMM  \\
        \midrule
        \midrule
        $R_{500}$ & \multirow{2}{*}{$877.8 \pm 46.2$} & \multirow{2}{*}{$780.9 \pm 19.8$} & \multirow{2}{*}{$810.1 \pm 41.9$}  \\
        $[\mathrm{kpc}]$ &  &  &  \\
        \midrule
        $\mathcal{D}_\mathrm{A}^2 \; Y_{500}$ & \multirow{2}{*}{$4.07 \pm 1.13$} & \multirow{2}{*}{--} & \multirow{2}{*}{$3.76 \pm 0.39$} \\
        $\big[10^{-5}\;\mathrm{Mpc}^2\big]$ &  &  &  \\
        \midrule
        $M_{500}$ & \multirow{2}{*}{$3.5 \pm 0.8$} & \multirow{2}{*}{$2.48 \pm 0.70$} & \multirow{2}{*}{$3.79 \pm 0.58$}  \\[3pt]
        $\big[10^{14}\;\mathrm{M}_\odot\big]$ &  &  &  \\
        \bottomrule
    \end{tabular}
    \end{center}
    \footnotesize\textbf{Notes.} The ACT values were taken from Table 8 of \act;\ see \S\ref{subsec:integ}.
    \label{tab:integ}
\end{table}

\section{Summary and conclusions}\label{sec:conclu}

We presented the first analysis of an individual cluster in the NIKA2 SZ Large Program sample with standard data quality.
This analysis faced several challenges.
The target, ACT-CL J0215.4+0030, is a cluster of low mass and high redshift, making it a faint target (less than 1 mJy/beam at its peak) and one of the most compact objects of the NIKA2 LPSZ sample.
The cluster was observed with NIKA2 for a time estimated sufficient to reach a significance of $3\sigmaup$ at $\theta_{500}$ on the surface brightness profile, which we fall slightly short of with $3\sigmaup$ at $0.7\times\theta_{500}$.
Moreover, the NIKA2 data are strongly contaminated by  point sources.
This contamination greatly affects the SZ signal because of the small angular size of the cluster, but also because of the strong fluxes of point sources.
The possibility of masking the point source contamination when no external data are available for spectrum fitting will be the subject of a future study.
Interferometric follow-ups using the Northern Extended Millimeter Array (NOEMA) are also ongoing. They allow directly measuring the point source contamination in the $uv$ plane.

Despite these challenges, we were able to extract the thermodynamical properties of the ICM.
The point source contamination was accounted for in a multiwavelength analysis and was used in a joint fit of the pressure distribution of the ICM and of the point source contamination.
This way of considering the contamination allowed us to fully take the uncertainty on the fluxes of point sources into account in our MCMC analysis.
The result of the joint fit gave us access to the pressure profile of the cluster.
This pressure profile was compared with the one obtained from the stand-alone analysis of XMM-\textit{Newton} data.
The results were found to be compatible.
The combination of our NIKA2 pressure profile and of the density profile obtained from
XMM-\textit{Newton} also allowed us to explore additional thermodynamical properties of the cluster.
Our conclusions are listed below.
\begin{itemize}
\item
The pressure profile of ACT-CL~J0215.4+0030 is found to be compatible with the profile of disturbed clusters found in \upp, but not with the universal profile from the same reference, indicating a disturbed core.
Other thermodynamical properties of the ICM (\eg,\ its core entropy) also indicate that the cluster has a disturbed core.

\item
The thermodynamical properties obtained from the combination of NIKA2 and XMM-\textit{Newton} are very competitive with those obtained using X-ray spectroscopy.
Because the exposure time needed to be able to extract spectroscopic information and therefore a temperature profile from X-ray observations is much longer than the time needed to be able to extract a density profile, and because this time increases steeply with redshift, the combination of X-rays and SZ is very well adapted for the measurement of thermodynamical properties of the ICM at high-$z$.

\item
The NIKA2 observations of this cluster allowed us to improve the precision of the cluster mass, both from previous knowledge from the ACT survey and the $Y_{500}-M_{500}$ scaling relation from \upp, and from X-ray only measurements.
\end{itemize}

Because this analysis studied the worst-case scenario for the NIKA2 SZ Large Program clusters, we can expect comparable or better precision for most of the clusters of the LPSZ. 
This is therefore a promising indication of our ability to use NIKA2 capabilities to achieve a precise calibration of the tools needed for cluster cosmology using SZ-detected catalogs of galaxy clusters.

\begin{acknowledgements}
   We would like to thank the IRAM staff for their support during the campaigns. 
   The NIKA dilution cryostat has been designed and built at the Institut N\'eel. In particular, we acknowledge the crucial contribution of the Cryogenics Group, and in particular Gregory Garde, Henri Rodenas, Jean Paul Leggeri, Philippe Camus. 
   This work has been partially funded by the Foundation Nanoscience Grenoble and the LabEx FOCUS ANR-11-LABX-0013. 
   This work is supported by the French National Research Agency under the contracts "MKIDS", "NIKA" and ANR-15-CE31-0017 and in the framework of the "Investissements d’avenir” program (ANR-15-IDEX-02). 
   This work has benefited from the support of the European Research Council Advanced Grants ORISTARS and M2C under the European Union's Seventh Framework Programme (Grant Agreement Nos. 291294 and 340519).
   F.R. acknowledges financial supports provided by NASA through SAO Award Number SV2-82023 issued by the Chandra X-Ray Observatory Center, which is operated by the Smithsonian Astrophysical Observatory for and on behalf of NASA under contract NAS8-03060.
   MDP acknowledges support from Sapienza Università di Roma through Progetti di Ricerca Medi 2019, prot. RM11916B7540DD8D.
   The results reported in this article are based on data obtained with XMM-Newton, an ESA science mission with instruments and contributions directly funded by ESA Member States and NASA. 
   This work was supported by CNES. 
   This research made use of \texttt{astropy}\footnote{http://www.astropy.org}, a community-developed core Python package for Astronomy \citep{astropy_collaboration_astropy:_2013,astropy_collaboration_astropy_2018}, as well as \texttt{scipy} \citep{jones_scipy_2001} and \texttt{matplotlib} \citep{hunter_matplotlib:_2007}.
\end{acknowledgements}

\bibliographystyle{bibtex/aa} 
\bibliography{NIKA2_ACT} 

\begin{appendix}
    \section{NIKA2 map residual noise and transfer function}\label{ap:noisetf} 
\begin{figure}[h]
        \begin{center}
        \includegraphics[width=\linewidth, trim={0cm 0cm 0cm 1cm}, clip]{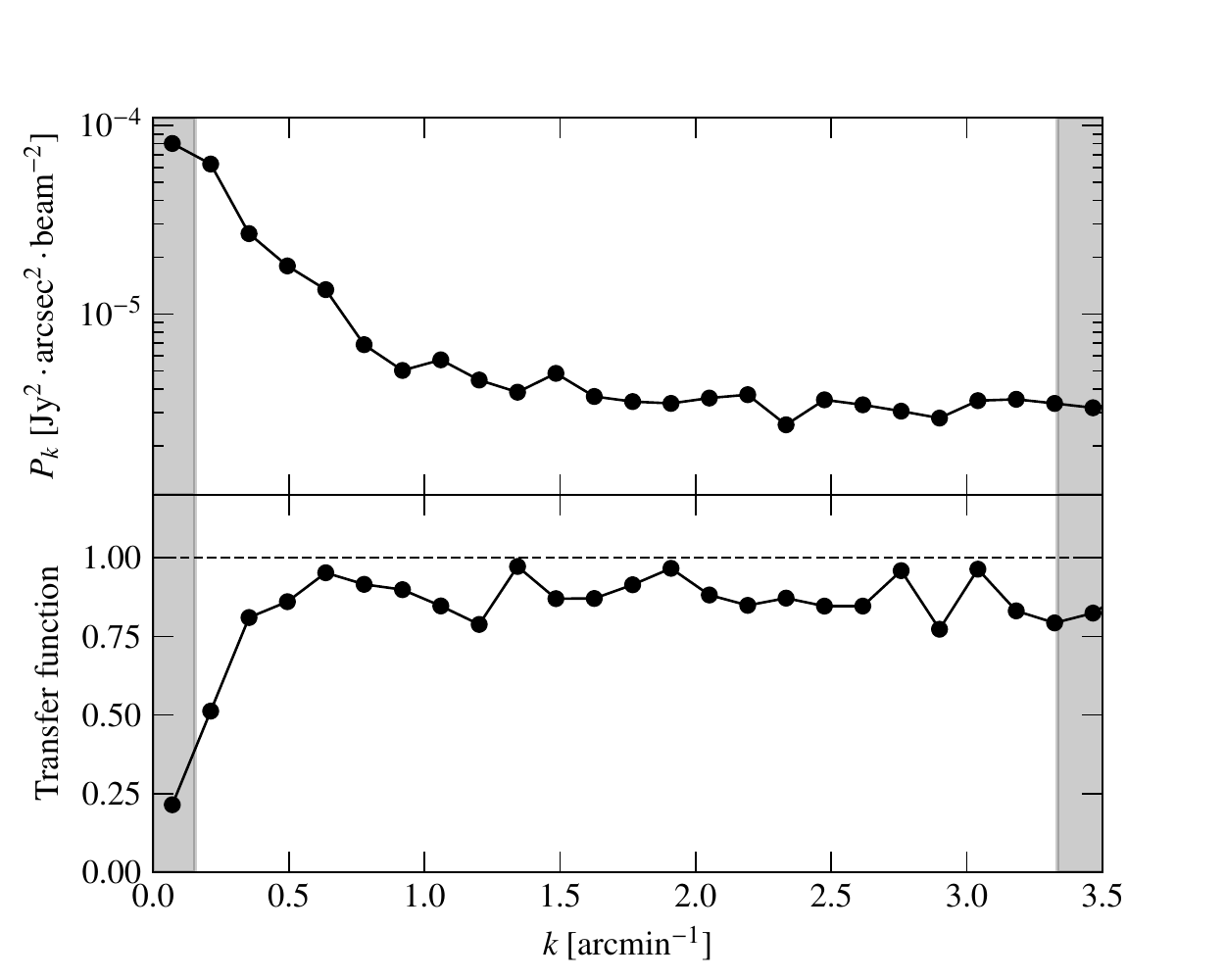}
        \caption{%
                Null-map power spectrum (\textit{top}) and transfer function (\textit{bottom}) of our NIKA2 150 GHz data. %
                The steepness of the power spectrum indicates large-scale correlated noise in our maps, which is accounted for in our analysis by computing a noise-covariance matrix. %
                The transfer function, evaluated on simulations, quantifies the filtering our data went through. %
                The angular scales are correctly recovered (with a transfer function above 80\%) for $k > 0.5\;{\rm arcmin^{-1}}$. %
                For each panel, the shaded gray regions represent the NIKA2 field of view (left) and instrumental FWHM (right).
            }
        \label{fig:tf_noise}
        \end{center}
    \end{figure}

    The noise levels in our NIKA2 maps and the filtering due to our data reduction process need to be known for our estimation of thermodynamical properties of the ICM.
    The power spectrum of the residual noise is estimated on a null map, which is often referred to as a \textit{\textup{jackknife}} map. 
    It is presented for the 150 GHz map in the top panel of Fig.~\ref{fig:tf_noise}.
    The power spectrum is steep, hence there is large-scale correlated noise in our maps.
    The pixel-by-pixel covariance matrix of the noise is therefore computed as the covariance of $10^5$ Monte Carlo noise realizations from this power spectrum.
    
    The filtering due to data processing was computed by applying the complete noise-removal procedure to the simulated signal.
    The transfer function for our data analysis procedure was computed as the ratio between the power spectra of the input simulated map and of the pipeline output.
    The input simulation was computed as a cluster model with an \upp\ pressure profile with the mass and redshift of ACT-CL~J0215.4+0030, plus a white-noise realization.
    By doing so, we ensured that all angular scales were represented in our input signal and that the transfer function was representative of the full signal filtering. 
    It is presented in the bottom panel of Fig.~\ref{fig:tf_noise}.
    For angular scales higher than $0.5\;{\rm arcmin^{-1}}$, it is flat and shows a filtering of less than 20\%.
    It was used as a filter in our forward-modeling approach, described in \S\ref{sec:panco}, so that our model undergoes the same filtering as the data it is compared to.
        More details can be found in previous NIKA2 (and NIKA) SZ papers, for example, \citet{ruppin_first_2018}, \citet{adam_pressure_2015} and  \citet{adam_high_2016}.
    
\end{appendix}

\end{document}